Category of Article: Methods and Protocols

# GA-guided mD-VcMD: A genetic-algorithm-guided method for multi-dimensional virtual-system coupled molecular dynamics

Junichi Higo[1], Ayumi Kusaka[2], Kota Kasahara[3], Narutoshi Kamiya[1], Ikuo Fukuda[1], Kentaro Mori[1,4], Yutaka Hata[1], and Yoshifumi Fukunishi[5]

*[1]Graduate School of Simulation Studies, University of Hyogo, 7-1-28 Minatojima Minami- machi, Chuo-ku, Kobe, Hyogo 650-0047, Japan*

*[2]Institute for Protein Research, Osaka University, 3-2 Yamada-oka, Suita, Osaka 565-0871, Japan*

*[3]College of Life Sciences, Ritsumeikan University, 1-1-1 Noji-higashi, Kusatsu, Shiga 525-8577, Japan*

*[4]National Institute of Technology, Maizuru College, 234 Shiroya, Maizuru, Kyoto, 625-8511 Japan*

*[5]Cellular and Molecular Biotechnology Research Institute, National Institute of Advanced Industrial Science and Technology (AIST), 2-3-26, Aomi, Koto-ku, Tokyo, 135-0064, Japan*

Corresponding author: Junichi Higo. Graduate School of Simulation Studies, University of Hyogo, 7-1-28 Minatojima Minami- machi, Chuo-ku, Kobe, Hyogo 650-0047, Japan. e-mail: higo@protein.osaka-u.ac.jp

Running title: Genetic-algorithm-guided enhanced sampling method

**Abstract**

We previously introduced a conformational sampling method, a multi-dimensional virtual-system coupled molecular dynamics (mD-VcMD), to enhance conformational sampling of a biomolecular system by computer simulations. Here, we present a new sampling method, subzone-based mD-VcMD, as an extension of mD-VcMD. Then, we further extend the subzone-based method using genetic algorithm (GA), and named it the GA-guided mD-VcMD. Because the conformational space of the biomolecular system is vast, a single simulation cannot sample the conformational space throughout. Then, iterative simulations are performed to increase the sampled region gradually. The new methods have the following advantages: (1) The methods are free from a production run: I.e., all snapshots from all iterations can be used for analyses. (2) The methods are free from fine tuning of a weight function (probability distribution function or potential of mean force). (3) A simple procedure is available to assign a thermodynamic weight to snapshots sampled in spite that the weight function is not used to proceed the iterative simulations. Thus, a canonical ensemble (i.e., a thermally equilibrated ensemble) is generated from the resultant snapshots. (4) If a poorly-sampled region emerges in sampling, selective sampling can be performed focusing on the poorly-sampled region without breaking the proportion of the canonical ensemble. A free-energy landscape of the biomolecular system is obtainable from the canonical ensemble.

**Introduction**

    Enhanced conformational sampling/generalized ensemble methods are useful to search a wide conformational space of a molecular system [1,2]. These methods enhance sampling along a reaction coordinate (RC) and are powerful to explore considerably stable states (major basins). On the other hand, these methods might overlook less-stable states (minor basins): Suppose that a major basin and a minor basin, whose positions are separated in the full-dimensional conformational space, overlap to each other in the low-dimensional RC axis. As a result, conformational trapping may occur, and the minor basins may be overlooked [3,4]. This oversight does not cause a serious problem when the minor basin is out of scope. However, an alternative approach is required when the minor basin acts as bridges among the major basins or have key features for exerting biophysical functions of the system.

    Adaptive umbrella sampling [5,6] has a possibility to escape such an oversight or trapping problem because one can set the RC so that the major and minor basins are discriminated along the RC axis. However, this method requires fine tuning of a weight function (potential of mean force (PMF) along the RC adopted). Practically, the difficulty of the fine tuning may result in a very long simulation (or increment of the number of iterative simulations) to sample the conformational space widely.

    To escape these difficulties, we introduced a multi-dimensional virtual-system coupled molecular dynamics (mD-VcMD) simulation [7,8], which is referred to as original mD-VcMD (or original method simply) in this paper. This method enhances conformational sampling with repeating iterative simulations in a multi-dimensional RC space. This method is free from fine tuning of the weight function because the weight function does not appear in the proceeding of the iterative simulations. We applied this method to a large and complicated system consisting of a ligand (endothelin-1), and its receptor (human endothelin type B receptor; one of the GPCR proteins), where the receptor was wrapped by cholesterols, embedded in an explicit membrane, and surrounded by an explicit solvent [8].

    On the other hand, we encountered another difficulty in the study: Once a poorly-sampled region emerged in the multi-dimensional RC space in an iterative simulation, this region might cause conformational trapping in a subsequent iteration. Thus, a more robust method was needed to proceed the iteration. In this paper, for the robust sampling, we introduce two methods with extending mD-VcMD: A subzone-based mD-VcMD method, which is an extension of mD-VcMD, and a GA-guided mD-VcMD method, which uses genetic algorithm (GA) to expand sampling to non-sampled RC regions.

Recently, we have applied the GA-guided method to molecular-binding systems, and showed that this method can provide a free-energy landscape consisting of the lowest free-energy complex (native-like complex), intermediate complexes (encounter complexes), and the unbound conformations [10].

**Methods and Protocols**

We introduce many terms and quantities in this paper and Supportive Information (SI). Table S1 of SI lists the positions where the terms and quantities are defined or explained in the paper. The original mD-VcMD method [7,8] produces the statistical weight $Q_{cano}(L)$ at each zone $L$ of the RC space (precise definition of $Q_{cano}(L)$ is given in SI). Sections 1–4 of SI are preparatory sections to understand the original method, and section 5 of SI explains the original method itself. The subzone-based and GA-guided mD-VcMD methods (or referred to as subzone-based and GA-guided methods, respectively) are extension of the original method to compute $Q_{cano}(L)$.

Once a set of $Q_{cano}(L)$ is given in a wide RC space, one can proceed the mD-VcMD simulation. Actual simulation protocol is presented in section 4 of SI, which are useful for all of the original, subzone-based and GA-guided methods.

The subzone-based and GA-guided methods are designed to treat a multiple-RC space as well as the original mD-VcMD method. In fact, the original method was applied to a complicated system using a seven-dimensional RC space [8]. On the other hand, to make explanation simple, we use a one-dimensional (1D), two-dimensional (2D), or three-dimensional (3D) RC space in this paper. One can increase the space dimensionality straightforwardly.

**1. Subzone-based method to compute $Q_{cano}(L)$**

**1.1 Introduction of subzone and subzone's snapshot count (SS count)**

Here, we present the subzone-based method to compute $Q_{cano}(L)$. Note that the sections below are based on sections 1-5 of SI. Figure 1A, which is substantially the same as Fig. S2C, is a portion of a 2D RC space, where four linked zones shown by differently colored frames overlap in an intersection (shaded region). The arrows indicate directions to shift the linked zones to remove the inter-zone overlaps. Figure 1B is an extended RC space resulted from the zone shifting, where the virtual states are numbered as $L_1, \ldots, L_9$ with an arbitrary order. The subscripts $i$ for $L_i$ are not those to specify linked virtual states. We use parentheses to specify the linked virtual states as $L_{(i)}$. Note that the extended RC space is introduced for convenience to make explanation simple. The zone-overlap exists as in Fig. 1A in the actual mD-VcMD simulation.

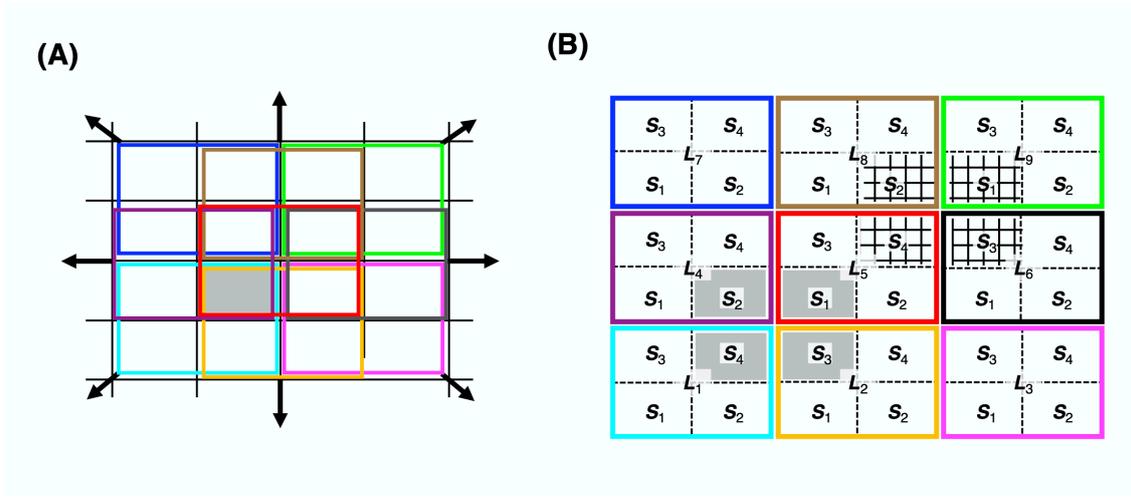

**Figure 1.** 2D RC space divided into zones. Panel (A) is equivalent to Fig. S2C in SI. Black arrows indicate directions to shift zones to remove inter-zone overlap (shaded region). (B) Extended RC space resulted from the zone shifts. $L_i$ ($i = 1, \ldots, 9$) and $S_j$ ($j = 1, \ldots, 4$) are the indices for zones and subzones, respectively. Shaded and checked subzones are mentioned in text.

Here we introduce *subzones*, which overlap perfectly on intersections in the original RC space. I.e., the shaded region in Fig. 1A corresponds to four shaded regions in Fig. 1B, and the subzones are these shaded regions in Fig. 1B. Similarly, four checked subzones in Fig. 1B overlap on another intersection (not shown in Fig. 1A). Once the subzones are introduced, any zone is constructed by subzones. For instance, in Fig. 1B, each zone $L_i$ consists of four subzones whose indices are expressed as $S_j$ ($j = 1, \ldots, 4$). In the multi-dimensional RC space, a zone is composed of $2^m$ subzones in general, where $m$ is the space dimensionality. Numbering of the subzones is arbitrary. Whereas $L$ specifies the zone position in the whole RC space, $S$ does the relative position of a subzone in each zone.

We express a subzone $S$ in zone $L$ as $[S, L]$. We denote the number of snapshots detected in the subzone during the iteration $M$ by $c^{[M]}(S, L)$, and normalize it within the zone $L$ as:

$$\frac{c^{[M]}(S,L)}{\sum_{S' \in L} c^{[M]}(S',L)} \to c^{[M]}(S, L). \tag{1}$$

This normalization results in: $\sum_{S \in L} c^{[M]}(S, L) = 1$. We refer to this quantity $c^{[M]}(S, L)$ as *subzone's snapshot count* (*SS count*) at $[S, L]$. The SS count for a non-sampled subzone in iteration $M$ is zero: $c^{[M]}(S, L) = 0$.

Here, we characterize zones in three types: If all the subzones in zone $\zeta_L$ are not sampled ($c^{[M]}(S, L) = 0$, $\forall S$ in $L$), this zone is called *empty zone* (*E zone*). If all the subzones in $L$ are sampled ($c^{[M]}(S, L) > 0$, $\forall S$ in $L$), this zone is referred to as *completely-sampled zone* (*CS zone*). Last, if some subzones are empty and the others are sampled in $L$, this zone is called an *incompletely sampled zone* (*IS zone*). Figure 2 exemplifies those zones assuming that the first, second, and third iterations are performed sequentially using a 2D RC space.

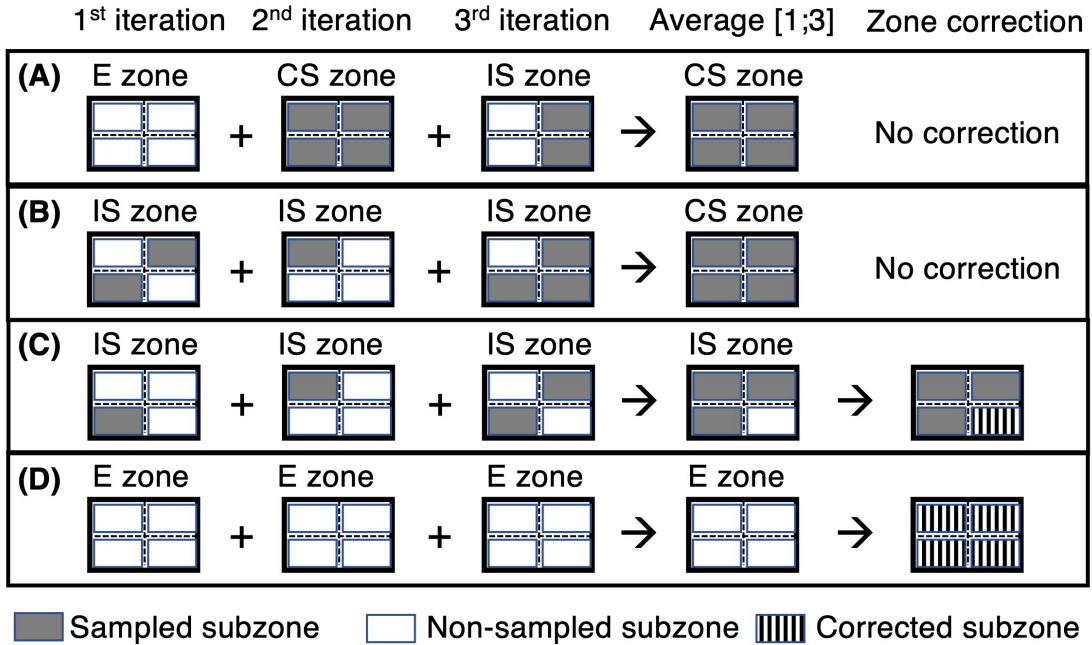

**Figure 2.** Time development of sampled and non-sampled subzones in a zone along three (1st, 2nd and 3rd) iterations. Four types of time development are exemplified from panel (A) to (D). "Average [1;3]" is $c^{[1;M]}(S,L)$ (Eq. 5) over the three iterations ($M = 3$). Shaded and white subzones are sampled and non-sampled ones, respectively. Checked subzones for "Zone correction" are corrected ones by Eq. 7 (procedure 2) for panel (C) and by Eq. 8 (procedure 3) for panel (D).

Next, we introduce two delta functions $\delta$ and $D$:

$$\delta^{[M]}(S,L) = \begin{cases} 1 & (\text{if } c^{[M]}(S,L) > 0) \\ 0 & (\text{if } c^{[M]}(S,L) = 0) \end{cases} \quad (2)$$

and

$$D^{[M]}(L) = \begin{cases} 1 & (\text{if } \zeta_L \text{ is CS zone in iteration } M) \\ 0 & (\text{otherwize}) \end{cases}. \quad (3)$$

Then, we average the SS count $c^{[I]}(S,L)$ ($I = 1, \ldots, M$) at each $[S,L]$ over the first to $M$-th iterations with the following procedures.

**Procedure 1:** If $\zeta_L$ is a CS zone in one or more iterations (Fig. 2A), the average of SS count is defined as:

$$c^{[1;M]}(S,L) = \frac{\sum_{I=1}^{M} D^{[I]}(L)\, c^{[I]}(S,L)}{\sum_{I=1}^{M} D^{[I]}(L)}. \quad (4)$$

IS and E zones are eliminated from the averaged SS counts. The superscript $[1;M]$ indicates that the SS counts are computed using an iteration windows from the first to $M$-th iterations.

**Procedure 2:** If $\zeta_L$ is an IS or E zone in all iterations and a CS zone is not involved, average is calculated as:

$$c^{[1;M]}(S,L) = \frac{\sum_{I=1}^{M} c^{[I]}(S,L)}{\sum_{I=1}^{M} \delta^{[I]}(L,S)}. \quad (5)$$

Note that the resultant SS counts do not vary when $c^{[I]}(S,L)$ is replaced by $\delta^{[I]}(S,L)c^{[I]}(S,L)$ in Eq. 5. A CS zone can be resulted from combining IS zones as shown in Fig. 2B.

Contrarily, in Fig. 2C, some non-sampled subzones (say $[S', L]$) may remain after averaging, which results in an IS zone for the zone and $c^{[1;M]}(S', L) = 0$. Then, we correct the zero SS counts as follows: First, we average non-zero SS counts in $L$:

$$\bar{c}(L) = \frac{\sum_{S \in L} c^{[1;M]}(S,L)}{\sum_{S \in L} \delta^{[1;M]}(S,L)}, \quad (6)$$

and replace the zero SS count by $\bar{c}(L)$:

$$c^{[1;M]}(S', L) = \bar{c}(L) \quad \text{(for non} - \text{sampled subzons)} \quad (7)$$

See Fig. 2C for an example for this case. Finally, all subzones in the IS zone have a non-zero value: $c^{(1;M)}(S, L) > 0 \ (\forall S \text{ in } L)$.

**Procedure 3:** All of the SS counts are zero if $\zeta_L$ is an E zone in all the iterations. Then, we correct $c^{(1;M)}(S, L)$ simply as:

$$c^{[1;M]}(S, L) = 1 \quad \text{(if } \zeta_L \text{ is E zone in all iterations)}. \quad (8)$$

See Fig. 2D for procedure 3.

**1.2 Computation of $Q_{cano}(L)$ from SS counts**

In this section, we present a method to calculate $Q_{cano}^{[M]}(L)$ by fitting SS counts $c^{[1;M]}(S, L)$ among neighboring subzones. The idea of subzone fitting is originated from our previous study [10]. Figure 5B of Ref. 10 indicates that fitting of fractions (short colored curves in the figure) of a canonical distribution produces the full distribution (black-line curve). Although the fitting method in Ref. 10 is function-fitting (not subzone-fitting), the logic is the same between Ref. 10 and the current study.

Suppose that there are $n_{is}$ intersections in the entire RC space, and that the intersections are numbered in an arbitrary order. If a couple of subzones overlap to each other in the RC space, these subzones are called a *subzone pair*. Here, the number of subzone pairs originated from the $j$-th intersection ($j = 1, ..., n_{is}$) is denoted by $n_{pair}(j)$: $n_{pair}(j) = n_{link}(n_{link} - 1)/2$ and $n_{link}$ is a function of $j$ ($n_{link} = n_{link}(j)$) because $n_{link}$ depends on the position of the $j$-th intersection in the RC space. Figure 3 is a 2D RC space presented in the extension form. Four subzones indicated by red circles are originated from an intersection, and the subzones generate six subzone ($= 4 \times 3/2$) pairs, which are indicated by red-colored lines. Similarly, four subzones indicated by blue circles are originated from another intersection, and six blue-colored lines specify subzone pairs. Similar argument is possible for the mD RC space. Suppose the $k$-th subzone pair with respect to the $j$-th intersection ($k = 1, \ldots, n_{pair}(j)$). We denote the indices of the pairing subzones by $p_{k,j}^{(1)}$ and $p_{k,j}^{(2)}$, where superscripts (1) and (2) indicate simply that the two subzones are pairing. Similarly, we denote the zone indices (virtual states) regarding the subzone pairs by $q_{k,j}^{(1)}$ and $q_{k,j}^{(2)}$. Then, the pairing subzones in the entire RC space are expressed by $[S_{p_{k,j}^{(1)}}, L_{q_{k,j}^{(1)}}]$ and $[S_{p_{k,j}^{(2)}}, L_{q_{k,j}^{(2)}}]$. Any of the pairing subzones in the RC space can be specified with varying $k$ and $j$.

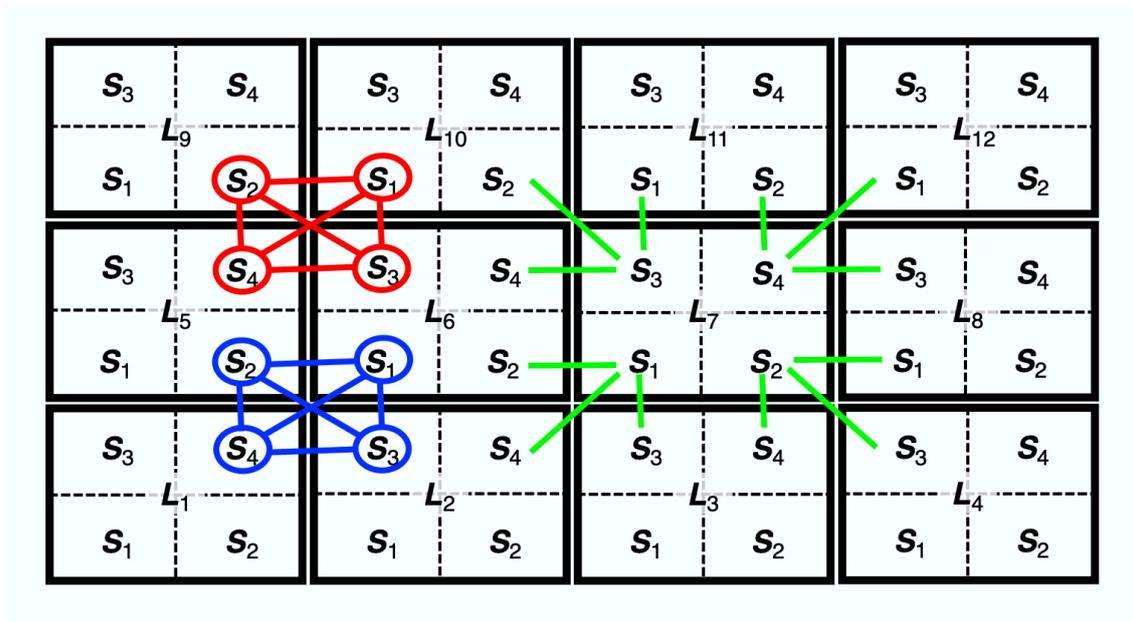

**Figure 3.** 2D RC space presented in extended form as Fig. 1B. Four subzones in red-colored circles overlap to a single intersection, and six red-colored lines indicate six subzone pairs. Similarly, four subzones in blue-colored circles overlap to another intersection with six subzone pairs indicated by blue-colored lines. Green-colored lines indicate subzone pairs that are related to a zone $\zeta_{L_7}$, which is mentioned in text.

Next, we define a *modulated SS count*, $\rho$, by multiplying a factor $\sigma_L$ to $c^{[1;M]}(\boldsymbol{S}, \boldsymbol{L})$:

$$\rho^{[M]}(\boldsymbol{S}, \boldsymbol{L}; \sigma_L) = \sigma_L c^{[1;M]}(\boldsymbol{S}, \boldsymbol{L}), \tag{9}$$

where $\sigma_L$ depends on $\boldsymbol{L}$. Last, we define an objective function $F$ as:

$$F(\boldsymbol{\sigma}) = \sum_{j=1}^{n_{is}} \sum_{k=1}^{n_{pair}(j)} (f_{j,k} - 1), \tag{10}$$

where

$$f_{j,k} = \frac{\max\left[\rho^{[M]}\left(\boldsymbol{S}_{p_{k,j}^{(1)}}, \boldsymbol{L}_{q_{k,j}^{(1)}}; \sigma_{L_{q_{k,j}^{(1)}}}\right), \rho^{[M]}\left(\boldsymbol{S}_{p_{k,j}^{(2)}}, \boldsymbol{L}_{q_{k,j}^{(2)}}; \sigma_{L_{q_{k,j}^{(2)}}}\right)\right]}{\min\left[\rho^{[M]}\left(\boldsymbol{S}_{p_{k,j}^{(1)}}, \boldsymbol{L}_{q_{k,j}^{(1)}}; \sigma_{L_{q_{k,j}^{(1)}}}\right), \rho^{[M]}\left(\boldsymbol{S}_{p_{k,j}^{(2)}}, \boldsymbol{L}_{q_{k,j}^{(2)}}; \sigma_{L_{q_{k,j}^{(2)}}}\right)\right]}. \tag{11}$$

The multiplication factors are packed in a vector form: $\boldsymbol{\sigma} = [\sigma_{L_1}, \sigma_{L_2}, \cdots]$. Equation 11 is invariant in exchange of superscripts (1) and (2). The double summations in Eq. 11 moves over all of the subzone pairs in the entire RC space. The "−1" in Eq. 10 is introduced to set $F$ to zero when $\rho^{[M]}(\boldsymbol{S}_{p_{k,j}^{(1)}}, \boldsymbol{L}_{q_{k,j}^{(1)}}; \sigma_{L_{q_{k,j}^{(1)}}}) = \rho^{[M]}(\boldsymbol{S}_{p_{k,j}^{(2)}}, \boldsymbol{L}_{q_{k,j}^{(2)}}; \sigma_{L_{q_{k,j}^{(2)}}})$.

Then, we minimize $F$ by modulating $\boldsymbol{\sigma}$, and refer to the resultant $\boldsymbol{\sigma}$ providing the minimized $F$ as $\boldsymbol{\sigma}^{min} = [\sigma_{L_1}^{min}, \sigma_{L_2}^{min}, \ldots]$. Finally, $Q_{cano}^{[M]}(\boldsymbol{L})$ is given as:

$$Q_{cano}^{[M]}(\boldsymbol{L}) = \sum_{\boldsymbol{S}_i \in \boldsymbol{L}} \rho^{[M]}(\boldsymbol{S}_i, \boldsymbol{L}; \sigma_L^{min}), \tag{12}$$

where the summation is taken over all subzones in the zone $\zeta_L$.

Minimization of $F$ corresponds to fitting the distribution of $c^{[1;M]}(\boldsymbol{S}, \boldsymbol{L})$ among neighboring zones by modulating $\boldsymbol{\sigma}$. This fitting is shown schematically in Fig.

4, which is presented in a 1D RC space to make the fitting procedure simple. Figure 4A plots values of the SS counts $c^{[1;M]}(S,L)$ before fitting, where the SS counts of two pairing subzones (i.e., two open circles in a broken-line rectangle) are not the same to each other. Figure 4B represents the modulated SS counts after fitting, where the two modulated SS counts pairing to each other (the filled circles in the rectangle) have the same value.

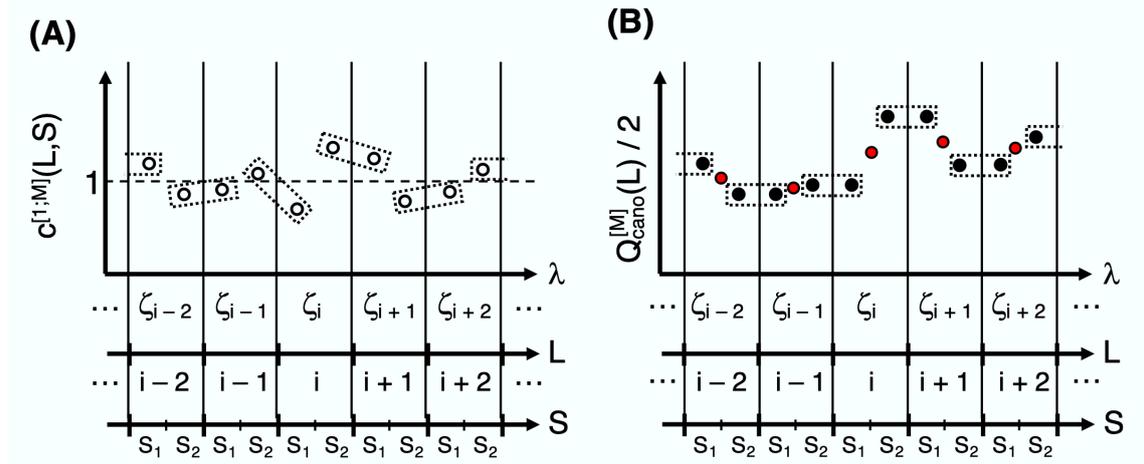

**Figure 4.** In both panels, the x-axis is presented by three measures: $\lambda$, $L$, and $S$, which are scalar because the RC space is 1D in this figure. (A) Open circles represent SS counts $c^{[1;M]}(L,S)$, and pairing subzones are connected by broken-line rectangles. Broken line indicates the level of $c^{[1;M]}(L,S) = 1$. (B) Filled circles represent modulated SS counts $\rho^{[M]}(S_k, L_j; \sigma_L^{min})$ and red circles are $Q_{cano}^{[M]}(L)/2 = \{\rho^{[M]}(S_1, L_k; \sigma_{L_k}^{min}) + \rho^{[M]}(S_2, L_k; \sigma_{L_k}^{min})\}/2$. Pairing subzones have the same modulated SS count.

In the 1D RC space, the objective function $F$ converges analytically to zero by the minimization. In a 2D or higher dimensional RC space, however, $F$ may not converge to zero and $\boldsymbol{\sigma}^{min}$ is not determined analytically. Thus, one may minimize $F$ by a Monte Carlo simulation with modulating $\boldsymbol{\sigma}$. Using the converged $\boldsymbol{\sigma}^{min}$, $Q_{cano}^{[M]}(\boldsymbol{L})$ is calculated (Eq. 12), and $J_{\boldsymbol{L}^{(i)}}$ is set as:

$$J_{\boldsymbol{L}_{(i)}}^{[M]} = \left[Q_{cano}^{[M]}(\boldsymbol{L}_{(i)}) \sum_{j=1}^{n_{link}} \frac{1}{Q_{cano}^{[M]}(\boldsymbol{L}_{(j)})}\right]^{-1}, \qquad (13)$$

which is equivalent to Eq. S7 in SI. Then the $(M + 1)$-th iteration is performed using $J_{L_{(i)}}^{[M]}$.

Remember that $Q_{cano}^{[M]}(L)$ is calculated using $Q_{cano}^{[M-1]}(L)$ in the original method (Eq. S8 of SI), and that the convergence check ($Q_{entire}^{[M]}(L) \approx const$) is required (see section 5 of SI and ref. 10). The subzone-based method does not require either $Q_{cano}^{[M-1]}$ nor the convergence check. This is an advantage of the subzone-based method (and also the GA method explained later).

Furthermore, we can assess the accuracy of the resultant $Q_{entire}^{[M]}(L)$ locally in each zone as follows: Given a zone $\zeta_L$, first, we list subzone pairs relating to the zone. See green-colored lines in Fig. 3, which illustrates subzone pairs related to a zone $\zeta_{L_7}$. We express the $i$-th subzone pair as $[S_{p_i^{(1)}}, L]$ and $[S_{p_i^{(2)}}, L_{q_i^{(2)}}]$. The superscript (1) indicates the subzone $[S_{p_i^{(1)}}, L]$ in $\zeta_L$, and (2) does the subzone $[S_{p_i^{(2)}}, L_{q_i^{(2)}}]$, which is not involved in $\zeta_L$. Then, a local function, $E_{local}(L)$, is defined as:

$$E_{local}(L) = \frac{1}{N_{pair}(L)} \sum_{i=1}^{N_{pair}(L)} \left\{ \frac{\max\left[\rho^{[M]}\left(S_{p_i^{(1)}}, L; \sigma_L^{min}\right), \rho^{[M]}\left(S_{p_i^{(2)}}, L_{q_i^{(2)}}; \sigma_{L_{q_i^{(2)}}}^{min}\right)\right]}{\min\left[\rho^{[M]}\left(S_{p_i^{(1)}}, L; \sigma_L^{min}\right), \rho^{[M]}\left(S_{p_i^{(2)}}, L_{q_i^{(2)}}; \sigma_{L_{q_i^{(2)}}}^{min}\right)\right]} - 1 \right\}, (14)$$

where $N_{pair}(L)$ is the number of subzone pairs related to the zone $\zeta_L$, and the summation is taken over all subzone pairs related to the zone $\zeta_L$. Equation 14 can be regarded as a part of Eq. 10.

If the subzone-based mD-VcMD simulation is long enough, the modulated SS counts fit well between the pairing subzones in theory, which results in $E_{local}(L) \to 0$ ($\forall L$). In an actual simulation, sampling accuracy may be uneven: Some RC regions may be sampled poorly. If a region is not important biophysically, one need not to sample this region further. Contrarily, if the region is important, the region should be sampled selectively. We present a simple method for selective sampling in section 3.

We can judge that a zone $\zeta_L$ is accurately sampled when $E_{local}(L) \leq E_{sh}$ is satisfied, where $E_{sh}$ is a threshold set by a user. In our actual sampling, we set $E_{sh} = 0.25$ [10].

## 2. GA to determine $c^{[1;M]}(S, L)$ for IS and E zones

Section 1.1 (procedures 1–3) proposed the method to fill SS counts $c^{[1;M]}(S, L)$ for all of the CS, IS and E zones from the $M$-th iteration. Then, we may proceed to the next iteration (iteration $M + 1$) using the SS counts. However, these procedures do not presume the SS counts for IS and E zones. Here, we introduce the GA-guided method to fill $c^{[1;M]}(S, L)$ in the IS and E zones.

### 2.1 GA database

For preparation, we generate a database (called *GA database*) used for the GA procedure as follows: First, we consider a compact block composed of CS zones:

$$\{\zeta_L: L = [i + \Delta i, j + \Delta j, k + \Delta k, \ldots]; \Delta i, \Delta j, \Delta k, \ldots = -1, 0, +1]\}. \quad (15)$$

This block is called a *GA block*. We refer to the zone at the center of the block ($\Delta i = \Delta j = \Delta k = \cdots = 0$) as a *central GA zone*. Figures 5A and 5B illustrate the blocks in a 2D and 3D RC spaces, respectively. In the 2D RC space, GA blocks are classified into three types: inside, side, and corner GA blocks, depending on the position of the central GA zone in the RC space. In the 3D RC space, four types of block are possible: inside, face, side, or corner GA blocks. If a block is not an inside one, the variations of $\Delta i$, $\Delta j$, $\Delta k$, ... in Eq. 15 may range from $0$ to $+1$ or from $-1$ to $0$. In a higher dimensional RC space, we should classify the blocks into more than four types. When the $M$-th iteration has done, we have SS counts from various iteration windows: $c^{[1;1]}(S, L)$, $c^{[1;2]}(S, L)$, ..., $c^{[1;M]}(S, L)$, which are computed by Eqs. 4 and 5. Then, the GA database consists of GA blocks from the various iteration windows.

Figure 6A is an example of GA blocks in the 2D RC space. We designate the GA blocks as $GA_j$ ($j = 1, \ldots, 6$), where $j$ is set in an arbitrary order. The GA database may involve GA blocks that are taken from the same area of the RC space but from different iteration windows. We regard those GA blocks as different ones in the GA database. Thus, the number of GA blocks in the GA database increases with proceeding the iteration.

Each zone in $GA_i$ is expressed by its relative position to the central GA zone, and the relative position is expressed by a relative index $\Lambda$: $\Lambda = [\Lambda^{(\alpha)}, \Lambda^{(\beta)}, \Lambda^{(\gamma)}, \ldots]$ and $\Lambda^{(h)} = -1, 0, 1$ ($h = \alpha, \beta, \gamma, \ldots$) if the block $GA_i$ is an inside block. For a non-inside block, $\Lambda_j$ may range as $\Lambda^{(h)} = -1, 0$ or $\Lambda^{(h)} = 0, -1$. Remember that a subzone $\mathbf{S}$ in a zone $\mathbf{\Lambda}$ is expressed by its relative position in the zone (Fig. 1B). Then, we specify the subzone by $[\mathbf{S}, \mathbf{\Lambda}]$, and the SS count for the subzone $[\mathbf{S}, \mathbf{\Lambda}]$ in $GA_i$ is expressed by $c^{GA_i}(\mathbf{S}, \mathbf{\Lambda})$.

## 2.2 Recovery zone

Next, we proceed to estimation of SS counts for IS and E zones without using procedures 2 and 3 in section 1.1. After the $M$-th iteration, we pick an IS or E zone. Assume that the picked zone is surrounded by $N_{surr}$ zones. For instance, in the 2D RC space, $N_{surr} = 8, 5$ or $3$ for an inside, side, or corner zone, respectively (see Fig. 5A), and for the 3D RC zone, $N_{surr} = 26, 17, 11$, or $7$ for an inside, face, side, or corner zone, respectively (Fig. 5B). Then, if the picked zone is surrounded by more than $N_{surr}/2$ CS zones, this picked zone is referred to as a *recovery zone* (*REC zone*). Figure 6B exemplifies four REC zones: REC zones 1, 2, and 3 are surrounded by six, six, and five CS zones of eight zones ($N_{surr} = 8$), respectively, and REC zone 4 is surrounded by four CS zones of five zones ($N_{surr} = 5$). If a zone is judged as a REC zone, a block consisting of the REC zone and zones surrounding the REC zone is called a *recovery block* (*REC block*). The $i$-th REC block is denoted by $REC_i$. As well as GA blocks, REC blocks are classified into some types such as inside, corner, ….

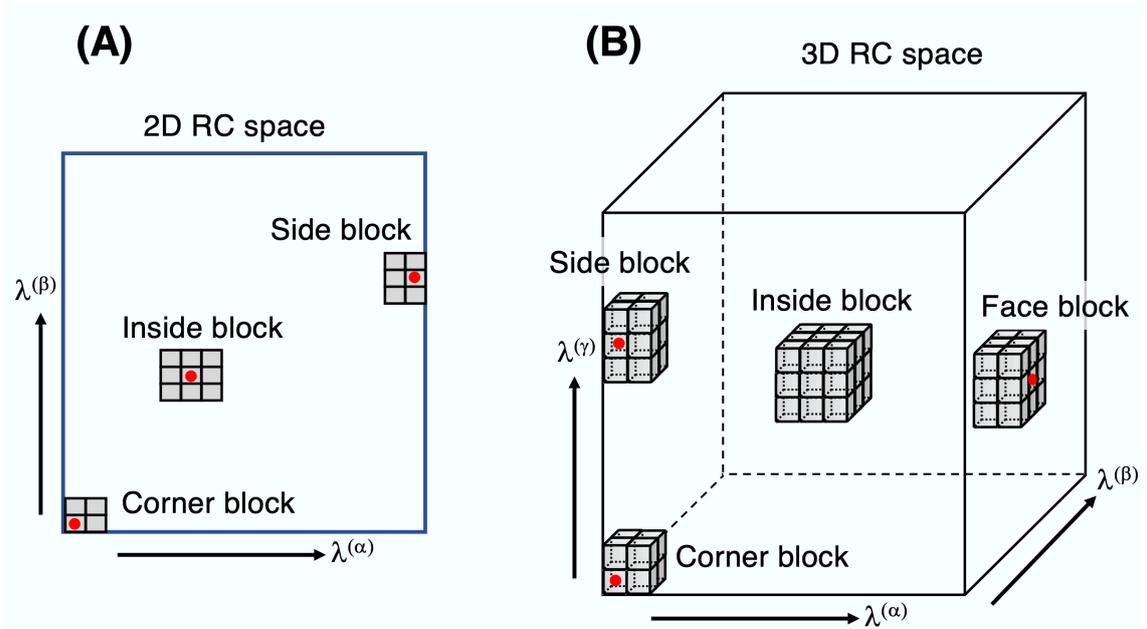

**Figure 5.** (A) GA blocks in 2D RC space. A GA block consists of CS zones, and is classified into three depending on the position of the central GA zone (marked by a red-colored circle) in the RC space: A block, whose central GA zone is at the side or corner of the RC space, is called a side or corner GA block, respectively. If the central GA zone is inside the RC space, the block is called an inside GA block. (B) In 3D RC space, four types of blocks are possible: inside, face, side, and corner. The central GA zone for the inside GA block is hidden. Although this figure is presented for blocks in GA database, the same block classification is possible for recovery blocks and blocks in GA generation $Gen_k$.

We note that more REC zones can be taken than those shown in Fig. 6A, and that the REC zones can overlap in the RC space: $REC_1$ and $REC_2$ overlap, and $REC_2$ and $REC_3$ do in the figure. We also note that the SS counts in a REC block, $c^{[1;M]}(S, L)$, are computed only from the iteration window $[1; M]$ (Eqs. 4 and 5), in contrast to those in a GA block, which are computed from various iteration windows $[1; 1], \dots, [1; M]$.

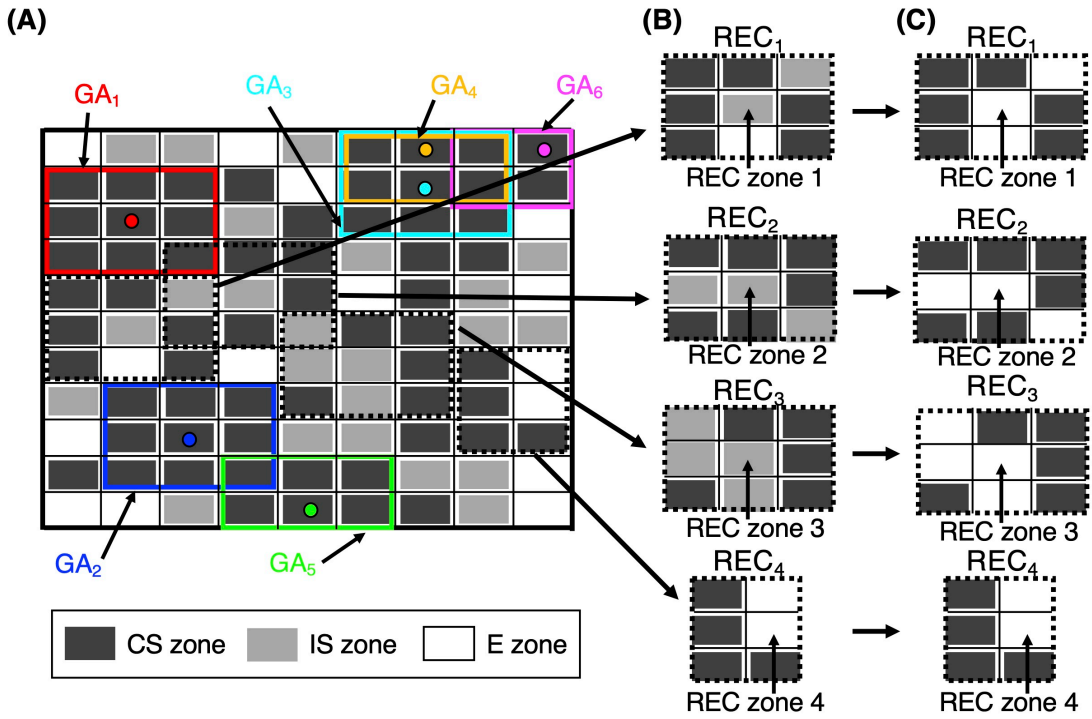

**Figure 6.** (A) GA blocks and recovery blocks in 2D RC space. Six GA blocks $GA_1, \dots, GA_6$ are shown by frames of different colors, and its central GA zone is marked by a circle with the same color. CS, IS, and E zones in GA blocks are shown by rectangles of different tones. Broken-line frames represent recovery blocks. (B) Four recovery blocks ($REC_1, \dots, REC_4$) and their recovery zones (REC zone 1, …, 4) picked from panel (A). Although more recovery zones are possible in panel (A), we do not discuss them. (C) Reset recovery blocks $REC_1, \dots,$ and $REC_4$.

**2.3 Resetting REC block**

We use indices $[S, \Lambda]$ to specify subzones in a REC block, as well as those in a GA block, and refer to a SS count $c^{[1;M]}(S, \Lambda)$ in $REC_i$ as $c^{REC_i}(S, \Lambda)$. Remember that the REC zone in all REC blocks is IS or E zone (Fig. 6B). Now we reset the SS counts $c^{REC_i}(S, \Lambda)$ of $REC_i$ to zero: $c^{REC_i}(S, \Lambda) \to 0$ ($\forall S$; $\Lambda = [0,0,0]$), which means that the REC zones become E zones (Fig. 6C). The SS counts in this reset REC zone is estimated by the GA procedure below.

**2.4 Similarity of spatial patterns of SS counts between of $REC_i$ and $GA_j$**

To assess similarity of the SS counts between blocks $REC_i$ and $GA_j$, we introduce a correlation coefficient $c_{cor}(REC_i, GA_j)$:

$$c_{cor}(REC_i, GA_j) = \frac{<REC_i \times GA_j> - <REC_i><GA_j>}{SD_{REC_i} \times SD_{GA_j}}, \quad (16)$$

where

$$<REC_i> = \frac{1}{N_{S\Lambda}} \sum_{\Lambda \in REC_i} \sum_{S \in \Lambda} c^{REC_i}(S, \Lambda), \quad (17)$$

$$<GA_j> = \frac{1}{N_{S\Lambda}} \sum_{\Lambda \in GA_j} \sum_{S \in \Lambda} \delta_{REC_i}(S, \Lambda) c^{GA_j}(S, \Lambda), \quad (18)$$

$$SD_{REC_i} = \left[ \frac{1}{N_{S\Lambda}} \sum_{\Lambda \in REC_i} \sum_{S \in \Lambda} c^{REC_i}(\Lambda, S)^2 - <REC_i>^2 \right]^{1/2}, \quad (19)$$

$$SD_{GA_j} = \left[ \frac{1}{N_{S\Lambda}} \sum_{\Lambda \in GA_j} \sum_{S \in \Lambda} \delta_{REC_i}(S, \Lambda) c^{GA_j}(S, \Lambda) - <GA_i>^2 \right]^{1/2}, \quad (20)$$

$$<REC_i \times GA_j> = \frac{1}{N_{S\Lambda}} \sum_{\Lambda \in REC_i} \sum_{S \in \Lambda} c^{REC_i}(\Lambda, S) c^{GA_j}(\Lambda, S), \quad (21)$$

$$N_{S\Lambda} = \sum_{\Lambda \in REC_i} \sum_{S \in \Lambda} \delta_{REC_i}(S, \Lambda), \quad (22)$$

$$\delta_{REC_i}(S, \Lambda) = \begin{cases} 1 & (\text{If } c^{REC_i}(\Lambda, S) \neq 0) \\ 0 & (else) \end{cases}. \quad (23)$$

The double summations $\Sigma_{\Lambda \in REC_i}\Sigma_{S \in \Lambda}$ and $\Sigma_{\Lambda \in GA_j}\Sigma_{S \in \Lambda}$ are taken over all subzones in $REC_i$ and $GA_j$, respectively. In Eq. 21 the summation $\Sigma_{\Lambda \in REC_i}$ operates not only to $c^{REC_i}(\Lambda, S)$ but also to $c^{GA_j}(\Lambda, S)$: The same relative positions of $\Lambda$ in $GA_j$ and $REC_i$ are summed in a coordinated manner.

We convert $c_{cor}(REC_i, GA_j)$ to a score function $E_{score}(REC_i, GA_j)$ as:

$$E_{simi}(REC_i, GA_j) = \begin{cases} \frac{1}{c_{cor}(REC_i, GA_j)} - 1 & (\text{if } c_{cor}(REC_i, GA_j) > c_{sh}) \\ \frac{1}{c_{sh}} - 1 & (\text{if } c_{cor}(REC_i, GA_j) \leq c_{sh}) \end{cases}. \quad (24)$$

Because the delta function $\delta_{REC_i}(S, \Lambda)$ is used in $c_{cor}(REC_i, GA_j)$, a non-zero $c^{GA_j}(\Lambda, S)$ in $GA_j$ is eliminated from the computation of $E_{simi}$ if $c^{REC_i}(S, \Lambda) = 0$. Also note that the REC zone of $REC_i$ is always eliminated from $E_{simi}$ because $\delta_{REC_i}(S, \Lambda) = 0$ for $\Lambda = [0,0,0]$. The parameter $c_{sh}$ is introduced to avoid a negative value of $E_{simi}$ by a negative correlation $c_{cor}(REC_i, GA_j)$. Thus $c_{sh}$ can be set to a small positive value ($c_{sh} = 0.01$ for instance). Therefore, the smaller the $E_{simi}$, the more similar the spatial patterns of SS counts between $REC_i$ and $GA_j$.

If $E_{simi}$ is smaller than a certain value, one may replace the SS counts of the REC zone of $REC_i$ by those of the central GA zone of $GA_j$. However, we assess a quality of $GA_j$ in the next section to judge if the replacement is appropriate.

**2.5 Overlap of SS-count patterns among zones in a GA block**

To assess the quality of a GA block $GA_j$, we quantify the spatial overlap of SS count between overlapping subzones within the block. Consider two subzones $[S, \Lambda]$ and $[S', \Lambda']$ in $GA_j$. First, we introduce a $\delta$ function $\delta_{ov}(L, S; L', S')$, which is used as a flag to detect overlapping subzones:

$$\delta_{ov}(S, \Lambda; S', \Lambda') = \begin{cases} 1 & ([S, \Lambda] \text{ and } [S', \Lambda'] \text{ are overlapping}) \\ 0 & (else) \end{cases}. \quad (25)$$

Then, $\delta_{ov}(L, S; L', S') = 1$ for any subzone pair from the shaded ones in Fig. 1B.

Similar to Eq. 9, we scale the SS counts $c^{GA_j}(S, \Lambda)$ by a factor $\sigma_\Lambda$, and generate modulated SS counts as $\rho^{GA_j}(S, \Lambda; \sigma_L) = \sigma_\Lambda c^{GA_j}(S, \Lambda)$. Then, we define a function $E_{phys}$ as:

$$E_{phys}(GA_j) = \sum_{\Lambda \in GA_j}^{\Lambda \neq 0} \sum_{\Lambda' \in GA_j}^{\Lambda' \neq 0, \Lambda' \neq \Lambda} \delta_{ov}(S, \Lambda; S', \Lambda') P(\sigma_L, \sigma_{L'}), \qquad (26)$$

where

$$P(\sigma_L, \sigma_{L'}) = \frac{\max\left[\rho^{GA_j}(S, \Lambda; \sigma_L), \rho^{GA_j}(S, \Lambda; \sigma_{L'})\right]}{\min\left[\rho^{GA_j}(S, \Lambda; \sigma_L), \rho^{GA_j}(S, \Lambda; \sigma_{L'})\right]} - 1. \qquad (27)$$

The summation $\sum_{\Lambda' \in GA_j}^{\Lambda' \neq 0, \Lambda' \neq \Lambda}$ in Eq. 26 is taken pairs of different zones. The double summations are taken with eliminating zone pairs regarding the central GA zone ($\Lambda \neq 0$ and $\Lambda' \neq 0$). The number of zones in a GA block depends on the place where the block is taken in the RC space. The summation of Eq. 26 is taken over existing zones in the GA block. Although $F(\sigma)$ (Eq. 10) was defined for the whole RC space, $E_{phys}$ is defined for a single GA block.

Last, we minimize $E_{phys}$ with modulating $\{\sigma_L\}$ by a Monte-Carlo scheme. The $E_{phys}$ used in the next section is this minimized one.

**2.6 The first generation of GA**

To start the GA procedure, the first generation (denoted as $Gen_1$) should be set using the GA database. Now, we focus on $REC_i$, calculate $E_{simi}(REC_i, GA_j)$ and $E_{phys}(GA_j)$ for all of the GA blocks in the GA database, and define the score function:

$$E_{score}(REC_i, GA_j) = \omega_s E_{simi}(REC_i, GA_j) + \omega_p E_{phys}(GA_j), \qquad (28)$$

where $\omega_s$ and $\omega_p$ are weights set by user (1.0 and 2.0, respectively, in our study [9]). The smaller the $E_{score}$, the more similar the SS-count patterns of $GA_j$ to $REC_i$ and the better the overlap of SS counts among zones in $GA_j$.

Next, we sort the resultant scores $\{E_{score}(REC_i, GA_j), \forall j\}$ in an ascending order, and collect $N_{GA}^{all}$ GA blocks from the first score. Those $N_{GA}^{all}$ GA blocks construct the first generation, denoted as $Gen_1$.

The first generation is prepared for each of the REC blocks, and the size of a REC block depends on its place in the RC space as mentioned before. We express the size of $REC_i$ and $GA_j$ as $n_\alpha^{REC_i} \times n_\beta^{REC_i} \times n_\gamma^{REC_i} \times \cdots$ and $n_\alpha^{GA_j} \times n_\beta^{GA_j} \times n_\gamma^{GA_j} \times \cdots$, respectively. Then, the GA blocks used in $E_{score}$ should satisfy the following relations: $n_h^{REC_i} \le n_h^{GA_j}$ ($h = \alpha, \beta, \gamma, ...$).

**2.7 Generation update**

Suppose that we have the $k$-th GA generation, denoted as $Gen_k$, with respect to a block $REC_i$. The generation update from $Gen_k$ to $Gen_{k+1}$ is illustrated in Fig. 7. First, we pick $N_{GA}^{top}$ members from the first to $N_{GA}^{top}$-th scored members from $Gen_k$, and pass them to the next generation $Gen_{k+1}$ (see the red-framed rectangle in Fig. 7). Thus, $N_{GA}^{top}$ members in $Gen_{k+1}$ are taken from $Gen_k$ without modulation.

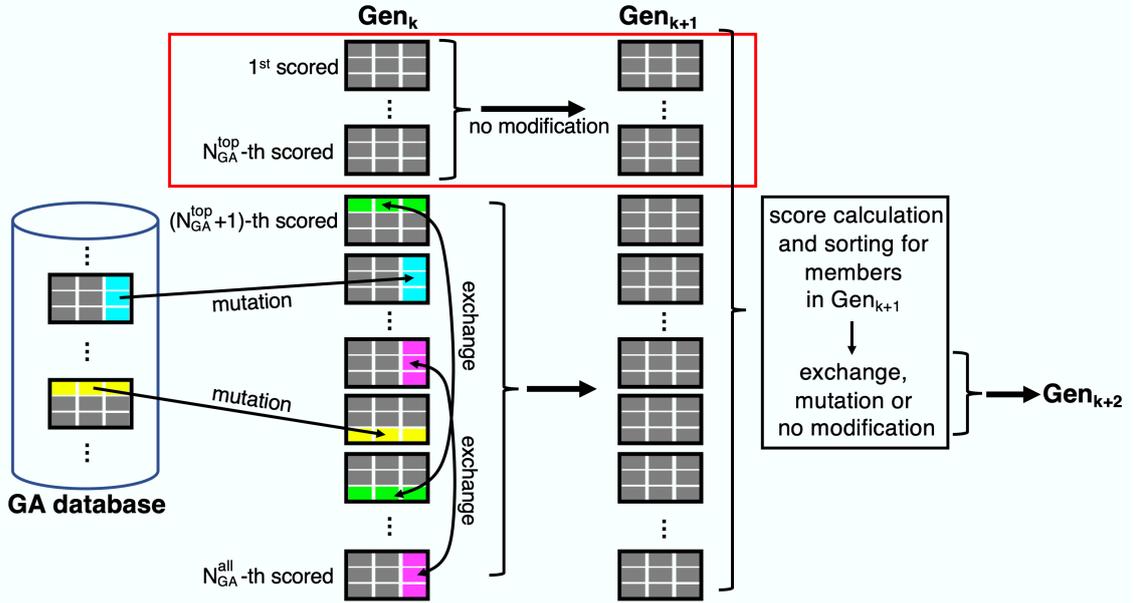

**Figure 7.** GA process to update the $k$-th GA generation $Gen_k$ to the $(k + 1)$-th one $Gen_{k+1}$. Details for the process are explained in text.

Next, we pick randomly $N_{GA}^{mut}$ members from $Gen_k$, which are denoted as $MEM_k^{(1)}, MEM_k^{(2)}, \ldots, MEM_k^{(N_{GA}^{mut})}$, and the same number of GA blocks from the GA database, denoted as $GA^{(1)}$, $GA^{(2)}$, …, $GA^{(N_{GA}^{mut})}$. Then, we replace a part of $MEM_k^{(l)}$ by a part of $GA^{(l)}$ ($l = 1, …, N_{GA}^{mut}$). This procedure corresponds to *mutation* in GA. In Fig. 7, the cyan-colored zones of a member in $Gen_k$ are replaced by the cyan-colored zones in a GA block. Similarly, the yellow-colored ones of a block in $Gen_k$ are replaced by the yellow-colored ones in a GA block. The mutated members are put in $Gen_{k+1}$.

Up to here, $N_{GA}^{top} + N_{GA}^{mut}$ generation members are set for $Gen_{k+1}$. Here, we pickup randomly $N_{GA}^{exc} (= N_{GA}^{all} - N_{GA}^{top} - N_{GA}^{mut})$ from $Gen_k$ and generate $N_{GA}^{exc}/2$ pairs of the picked members: $N_{GA}^{top}$ and $N_{GA}^{mut}$ should be set so that $N_{GA}^{exc}$ is an even number. We denote the $m$-th pair by $MEM_k^{(m)(1)}$-$MEM_k^{(m)(2)}$ ($m = 1, …, N_{GA}^{exc}/2$), where and superscripts (1) and (2) indicate that the two members are pairing. Then, a part of $MEM_k^{(m)(1)}$ and a part of $MEM_k^{(m)(2)}$ are exchanged. In Fig. 7, two magenta parts are exchanged mutually, and two green parts are done too. The exchanged members put in $Gen_{k+1}$. This procedure corresponds to *crossover* or *exchange* in GA. Up to here, $N_{GA}^{all}$ members are prepared for $Gen_{k+1}$. Details for mutation and exchange are explained later.

To generate $Gen_{k+2}$, we repeat similar procedures to $Gen_{k+1}$ as done for $Gen_k$: (1) Calculate the score function between each REC block (say $REC_i$) and all of the $N_{GA}^{all}$ members in $Gen_{k+1}$ (not all blocks in the GA database):

$$E_{score}(REC_i, MEM_{j,k+1}) = w_s E_{simi}(REC_i, MEM_{j,k+1}) + w_p E_{phys}(MEM_{j,k+1}), (29)$$

where $MEM_{j,k+1}$ is the $j$-th member ($j = 1, …, N_{GA}^{all}$) in $Gen_{k+1}$. (2) The first to $N_{GA}^{top}$ scored members of $Gen_{k+1}$ are passed to the next generation $Gen_{k+2}$. (3) Picking randomly $N_{GA}^{mut}$ members from $Gen_{k+1}$ and $N_{GA}^{mut}$ GA blocks from the GA database, mutations are executed. The generated $N_{GA}^{mut}$ members are passed to $Gen_{k+2}$. (4) Picking $N_{GA}^{exc}$ members randomly from $Gen_{k+1}$, generate $N_{GA}^{exc}/2$ pairs of members, and exchange is done in each pair. The resultant $N_{GA}^{exc}$ members are passed to $Gen_{k+2}$. Up to here, $N_{GA}^{all}$ members are prepared for $Gen_{k+2}$.

We repeat this cycle $N_{cycle}$ times ($N_{cycle}$ is set by user). Finally, the central zone of $REC_i$ is replaced by the central zone of the best-scored member of $Gen_{N_{cycle}}$. User may exit the cycle before reaching the $N_{cycle}$ cycle if $E_{simi}$ and $E_{phys}$ of a member become smaller than thresholds determined by user.

## 2.8 Multiple GA procedure

Figure 8A is a part of 2D RC space, where the magenta-colored circles are REC zones. The SS counts for these REC zones are estimated by GA (Fig. 8B). Then, one may calculate $Q_{cano}^{[M]}(L)$ and $J_{L_{(i)}}^{[M]}$ from the SS-counts, and proceed to the $(M+1)$-th iteration. In Fig. 8B, however, two zones indicated by cyan-colored circles emerge as REC zones anew. Thus, we can proceed the second GA procedure to estimate the cyan-colored zones, and Fig. 8C shows that the cyan-colored zones are determined. On the other hand, another REC zone (light-yellow-colored zone) appears in Fig. 8C. Thus, we can multiply the GA procedure. When a new REC zone does not appear anymore, we quit the procedure, and proceed to the $(M+1)$-th iteration.

Figure S3 of DI illustrates the procedures for the GA-guided mD-VcMD method as a flow chart.

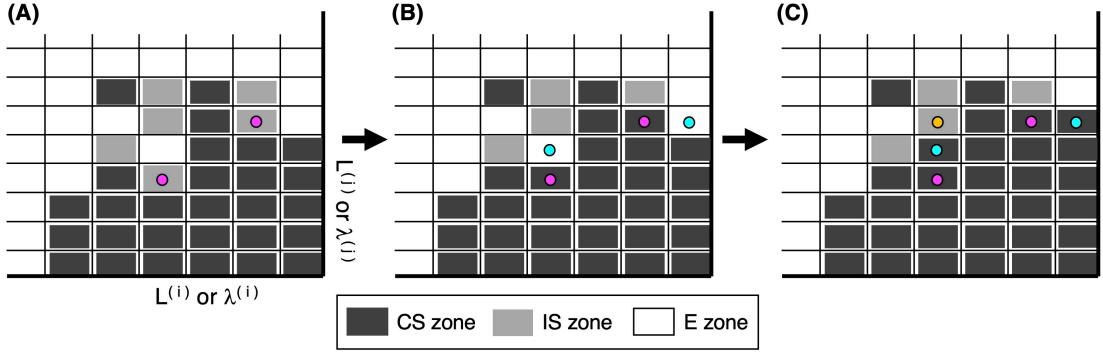

**Figure 8.** Part of 2D RC space, whose axes are $L^{(i)}$ (or $\lambda^{(i)}$) and $L^{(j)}$ (or $\lambda^{(j)}$). Three zone types are represented by different tones. Colored circles are REC zones mentioned in text.

## 2.9. Mutation and exchange

We explain details for mutation and exchange. In mutation, a pair of $MEM_k^{(i)}$ ($i = 1, \ldots, N_{GA}^{mut}$) and a GA block $GA^{(i)}$ is treated, and in exchange, a pair of $MEM_k^{(m)(1)}$ ($m = 1, \ldots, N_{GA}^{exc}/2$) and $MEM_k^{(m)(2)}$ is treated. In either operation, we select a slice from $MEM_k^{(i)}$, $GA^{(i)}$, $MEM_k^{(m)(1)}$, and $MEM_k^{(m)(2)}$. Imagine slices in a block as:

$$\Lambda^{(h)} = \begin{cases} -1 \\ +1 \end{cases} \quad (h = \alpha, \beta, \gamma, \ldots). \tag{30}$$

Figure 9 is an image of slices in the 3D RC space to help understanding the slices. Each slice consists of 9 zones.

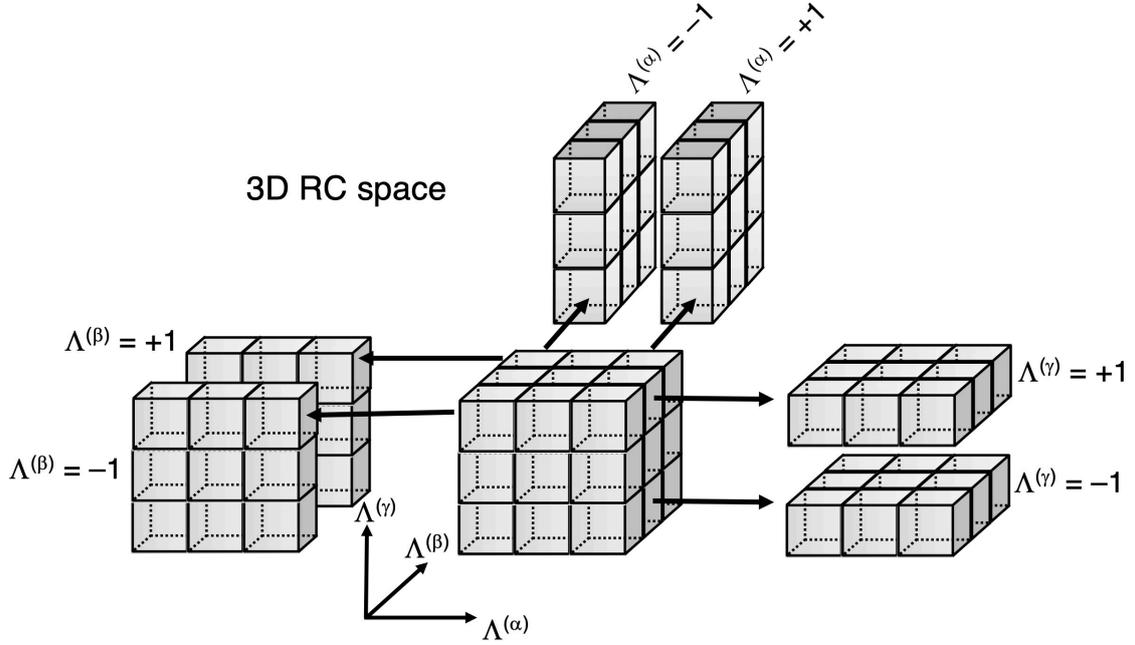

**Figure 9.** Image of slices in the 3D RC space.

In mutation, a slice, say $\Lambda^{(x)}$, is picked randomly from slices of $MEM_k^{(i)}$. Then, a slice parallel to $\Lambda^{(x)}$ is selected randomly from two slices $\Lambda^{(x)} = -1$ or $\Lambda^{(x)} = -1$ of $GA^{(i)}$. Then, the selected slice of $MEM_k^{(i)}$ is replaced by that of $GA^{(i)}$, which generates a new member for $Gen_{k+1}$.

In exchange, after selecting a slice randomly from slices of $MEM_k^{(m)(1)}$, a slice is selected randomly from two slices of $MEM_k^{(m)(2)}$ with the same manner for mutation. Then, the selected slices are exchanged mutually. This procedure generates two members for $Gen_{k+1}$.

We do not use a middle slice ($\Lambda^{(x)} = 0$) for the mutation and exchange. The GA procedure is used to estimate the SS counts of a REC zone, and the REC zone is involved in the middle slices of a REC block. To keep logical consistency between the REC zone and the member in $Gen_k$, we eliminated the middle slice.

We note a technical point: In mutation and exchange, the size of the operated slice of a REC block should be included in or the same as that of a block in $Gen_k$, and the size of the operated slice of the block in $Gen_k$ should be included in or the same as that of a GA block.

## 3. Selective sampling of poorly sampled RC region

Suppose that the $M$-th iteration has finished, and that an RC region was poorly sampled in all iterations. Then the accuracy of $Q_{cano}^{[M]}(L)$ for the poorly sampled RC region can be increased by a simple procedure: We pick conformations, which are in or near the poorly sampled RC region, from trajectories previously done. Then, the $(M + 1)$-th iteration is done starting from those conformations.

This selective sampling procedure does not work if $Q_{cano}^{[M]}(L)$ is calculated by the original method (Eq. S8 in SI) because this procedure increases $Q_{entire}^{[M]}(L)$ locally, and this local inflation of $Q_{entire}^{[M]}(L)$ breaks balance of $Q_{entire}^{[M]}$ resulting in errors of $Q_{cano}^{[M]}$ in a wide RC region.

Contrarily, in the subzone-based and GA-guided methods, the artificial increase of $Q_{entire}^{[M]}(L)$ vanishes by fitting $\rho^{[M]}(S, L; \sigma_L^{min})$ among overlapping zones. This is an advantage of the subzone-based and GA methods against the original method.

We introduced a selective sampling procedure in our previous study [8]. However, the selective sampling introduced here is different from that in ref. 8.

## 4. Thermodynamic weight assigned to sampled conformations

It is essentially important to assign a thermodynamic weight to each snapshot to analyze the resultant conformational ensemble. Given a zone $\zeta_L$, the thermodynamic weight assigned to $\zeta_L$ is $Q_{cano}(L)$. If the simulation is quit at the $M$-th iteration, $Q_{cano}(L) = Q_{cano}^{[M]}(L)$. In the original method, the weight assigned to the snapshot detected in $\zeta_L$ is proportional to $Q_{cano}(L)$ if $Q_{cano}(L) \approx const$ is satisfied in the simulation because the number of snapshots detected in zones is approximately even.

In the subzone-based and GA methods, however, sampling is flexible. For instance, by the selective sampling proposed above, the number of snapshots in zones may be considerably uneven, which means that the thermodynamic weight assigned to a snapshot is not equivalent to $Q_{cano}(L)$. We present a simple method for assigning weight as follows: Consider a snapshot, which is detected in $\zeta_L$, and suppose that this snapshot is the $i$-th one in the conformational ensemble. Then, the thermodynamic weight, $w_i$, assigned to the $i$-th snapshot is:

$$w_i = Q_{cano}(L)/n_{snap}^{[1;M]}(L), \tag{31}$$

where $n_{snap}^{[1;M]}(L)$ is the number of snapshots detected in $\zeta_L$ in all of the iterations. Although $w_i$ is dependent of $L$ in Eq. 31, once the weight is assigned, we can forget $L$. If a relation $n_{snap}^{[1;M]}(L) \approx const$ is satisfied, $w_i \approx Q_{cano}(L)$ substantially.

## Closing remarks

Benefits of the subzone-based and GA-guided methods are: (1) All snapshots from all iterations can be used for analysis. That is, no production run is required, or all simulations are production run. Contrarily, in the original mD-VcMD method or other enhanced sampling methods, the production run is mandatory usually, and only the production run is used for analysis. (2) To proceed sampling, no fine tuning of a weight function is used. This is an advantage against multicanonical or other adaptive umbrella sampling method. (3) The statistical weight $w_i$ (Eq. 31) is determined after sampling, and the method to compute $w_i$ is simple. (4) Statistical significance of a poorly-sampled region can be raised by selective sampling. In the original mD-VcMD method, the selective sampling accumulates local errors in $Q_{cano}(L)$, which results in a slow convergence.

GA-guided mD-VcMD is applicable to any complicated biomolecular system whose potential energy is defined for an MD simulation. We generated computer programs for the GA-guided mD-VcMD method, and applied it to molecular binding of a flexible medium-sized ligand (a peptide of about ten amino-acid residues long) to a receptor protein in an explicit solvent [10]. The resultant conformational ensemble produced a binding free-energy landscape where various structural clusters (the unbound state, fragile encounter complexes, and the most stable native-like complex) distributed. This showed that the GA-guided mD-VcMD simulation sampled the wide conformational space.

## Acknowledgements

We deeply appreciate Prof. Haruki Nakamura from Osaka Univ. for insightful comments. J. H. was supported by JSPS KAKENHI Grant No. 16K05517 and by the Development of core technologies for innovative drug development based upon IT from Japan Agency for Medical Research and Development (AMED). N. K. wad supported by JSPS KAKENHI Grant (Number JP20H03229). K. K. was also supported by JSPS KAKENHI Grant No. 16K18526. This work was supported by the HPCI System Research Project (Project IDs: hp190017, hp190018, hp190027, hp200063, hp200090, and hp200025). The simulations were performed on the TSUBAME3.0 supercomputers at the Tokyo Institute of Technology. It was performed in part under the Cooperative Research Program of the Institute for Protein Research, Osaka University, CR-19-05 and CR-20-05.

**References**

[1] Mitsutake, A., Sugita, Y. & Okamoto, Y. Generalized-ensemble algorithms for molecular simulations of biopolymers. *Biopolymers* **60**, 96–123 (2001). DOI: 10.1002/1097-0282(2001)60:2<96::AID-BIP1007>3.0.CO;2-F

[2] Higo, J., Ikebe, J., Kamiya, N. & Nakamura, H. Enhanced and effective conformational sampling of protein molecular systems for their free energy landscapes. *Biophysical Rev.* **4**, 27–44 (2012). DOI: 10.1007/s12551-011-0063-6

[3] Higo, J., Dasgupta, B., Mashimo, T., Kasahara, K., Fukunishi, Y. & Nakamura, H. Virtual-system-coupled adaptive umbrella sampling to compute free-energy landscape for flexible molecular docking. *J. Comput. Chem.* **36**, 1489–1501 (2015). DOI: 10.1002/jcc.23948

[4] Higo, J., Kasahara, K., Dasgupta, B. & Nakamura, H. Enhancement of canonical sampling by virtual-state transitions. *J. Chem. Phys.* **146**, 044104 (2017). DOI: 10.1063/1.4974087

[5] Paine, G. H. & Scheraga, H. A. Prediction of the native conformation of a polypeptide by a statistical-mechanical procedure. I. Backbone structure of enkephalin. *Biopolymers* **24**, 1391–1436 (1985). DOI: 10.1002/bip.360240802

[6] Mezei, M. Adaptive umbrella sampling: Self-consistent determination of the non-Boltzmann bias. *J. Comput. Phys.* **68**, 237–248 (1987). DOI: 10.1016/0021-9991(87)90054-4

[7] Hayami, T., Higo, J., Nakamura, H. & Kasahara, K. Multidimensional virtual-system coupled canonical molecular dynamics to compute free-energy landscapes of peptide multimer assembly. *J Comput. Chem.* **40**, 2453–2463 (2019). DOI: 10.1002/jcc.26020

[8] Higo, J., Kasahara, K., Wada, M., Dasgupta, B., Kamiya, N., Hayami, T., *et al.* Free-energy landscape of molecular interactions between endothelin 1 and human endothelin type B receptor: Fly-casting mechanism. *Protein Engineering, Design & Selection* (*PESD*) **32**, 297–308 (2019). DOI: 10.1093/protein/gzz029

[9] Higo, J., Kawabata, T., Kusaka, A., Kasahara, K., Kamiya, N., Fukuda, I., Mori, K., Hata, Y., Fukunishi, Y. & Nakamura, H. Molecular interaction mechanism of 14-3-3$\varepsilon$ protein with phosphorylated Myeloid leukemia factor 1 revealed by an enhanced conformational sampling. *bioRxiv* (2020). DOI: 10.1101/2020.05.24.113209

[10] Higo, J., Kasahara, K. & Nakamura, H. Multi-dimensional virtual system introduced to enhance canonical sampling. *J. Chem. Phys.* **147**, 134102 (2017). DOI: 10.1063/1.4986129

# GA-guided mD-VcMD: A genetic-algorithm-guided method for multi-dimensional virtual-system coupled molecular dynamics

Junichi Higo[1], Ayumi Kusaka[2], Kota Kasahara[3], Narutoshi Kamiya[1], Ikuo Fukuda[1], Kentaro Mori[1,4], Yutaka Hata[1], and Yoshifumi Fukunishi[5]

[1]*Graduate School of Simulation Studies, University of Hyogo, 7-1-28 Minatojima Minami- machi, Chuo-ku, Kobe, Hyogo 650-0047, Japan*

[2]*Institute for Protein Research, Osaka University, 3-2 Yamada-oka, Suita, Osaka 565-0871, Japan*

[3]*College of Life Sciences, Ritsumeikan University, 1-1-1 Noji-higashi, Kusatsu, Shiga 525-8577, Japan*

[4]*National Institute of Technology, Maizuru College, 234 Shiroya, Maizuru, Kyoto, 625-8511 Japan*

[5]*Cellular and Molecular Biotechnology Research Institute, National Institute of Advanced Industrial Science and Technology (AIST), 2-3-26, Aomi, Koto-ku, Tokyo, 135-0064, Japan*

Corresponding author: Junichi Higo. Graduate School of Simulation Studies, University of Hyogo, 7-1-28 Minatojima Minami- machi, Chuo-ku, Kobe, Hyogo 650-0047, Japan. e-mail: higo@protein.osaka-u.ac.jp

Running title: Genetic-algorithm-guided enhanced sampling method

**Table S1. Terminology.**

This table indicates where technical terms, parameters, and functions are defined in the main text or SI.

---

| Term/quantity | Section |
|---|---|
| Reaction coordinate (RC) | Introduction of main text or section 1 of SI |
| Multi-dimensional (mD) RCs | Introduction of main text or section 2 of SI |
| Atom groups $G_h^A$ and $G_h^B$ ($h = \alpha, \beta, \gamma, ...$) | Section 1 of SI (see Fig. S1) |
| $\lambda^{(h)}$ ($h = \alpha, \beta, \gamma, ...$) | Section 1 of SI (see Fig. S1) |
| $R$: system's conformation | Section 2 of SI |
| $\lambda = [\lambda^{(\alpha)}(R), \lambda^{(\beta)}(R), \lambda^{(\gamma)}(R), ...]$ | Section 2 of SI (see Fig. S2C) |
| $\zeta_i^{(h)}$ ($h = \alpha, \beta, \gamma, ...; i = 1, ..., n_{vs}(h)$) | Section 2 of SI (see Fig. S2A and S2B) |
| $n_{vs}(h)$ | Section 2 of SI (see Fig. S2A) |
| $\Delta\lambda_i^{(h)}$ and $\Delta\lambda^{(h)}$ | Section 2 of SI (see Fig. S2A) |
| $[\zeta_i^{(h)}]_{min}$ and $[\zeta_i^{(h)}]_{max}$ | Section 2 of SI (see Fig. S2A and S2B) |
| $L^{(h)}$ and $L = [L^{(\alpha)}, L^{(\beta)}, L^{(\gamma)}, ...]$: zone index | Section 2 of SI (see Fig. S2C) |
| $\zeta_L = \zeta_{[L^{(\alpha)}, L^{(\beta)}, L^{(\gamma)}, ...]} = [\zeta_{L^{(\alpha)}}^{(\alpha)}, \zeta_{L^{(\beta)}}^{(\beta)}, \zeta_{L^{(\gamma)}}^{(\gamma)}, ...]$ | Section 2 of SI (see Fig. S2C, and Eq. S1) |
| $L$: virtual-state variable (or virtual-state) | Section 3 of SI (equivalent to zone index) |
| intersection | Section 3 of SI (see Fig. S2C) |
| $\zeta_{L_{(1)}}, \zeta_{L_{(2)}}, ...,$ and $\zeta_{L_{(n_{link})}}$: linked RC zones | Section 3 of SI (see Fig. S2C) |
| $L_{(1)}, ..., L_{(n_{link})}$: linked virtual states | Sections 3 and 4 of SI |
| $n_{link}$: number of linked virtual states (zones) | Section 3 of SI (see Fig. S2C) |
| $E_{entire}(R, L)$: potential energy | Section 3 of SI (see Eq. S2) |
| $E_R(R)$: original potential energy | Section 3 of SI (see Eq. S2) |
| $E_V(L)$ | Section 3 of SI (see Eq. S5) |
| $E_{RV}(\lambda(R), L)$ | Section 3 of SI (see also Eq. S3) |
| inter-virtual state transition (IVT) = inter-zone transition | Sections 3 and 4 of SI |
| $P_t(L \to L')$: formal IVT probability | Section 4 of SI (see Eq. S6) |
| $J_{L_{(i)}}$: optimal IVT probability | Section 4 of SI (see Eq. 7) |
| configurational motion (CFM) | Section 4 of SI |
| $Q_{entire}(L)$: virtual state-partitioned probability | Section 4 of SI |
| $Q_{cano}(L)$: virtual state-partitioned canonical probability | Section 4 of SI |
| $S_j$: subzone index | Section 1.1 of main text (see Fig. 1B) |
| $c^{[M]}(S, L)$: subzone's snapshot count (SS count) | Section 1.1 of main text |
| empty zone (E zone) | Section 1.1 of main text |
| completely-sampled zone (CS zone) | Section 1.1 of main text |
| incompletely sampled zone (IS zone) | Section 1.1 of main text |
| $n_{is}$ | Section 1.2 of main text |
| subzone pair | Section 1.2 of main text |
| $n_{pair}(j)$ | Section 1.2 of main text |
| modulated SS count | Section 1.2 of main text |
| $p_{k,j}^{(1)}$ and $p_{k,j}^{(2)}$: pairing subzone indices | Section 1.2 of main text |
| $q_{k,j}^{(1)}$ and $q_{k,j}^{(2)}$: pairing zone indices | Section 1.2 of main text |
| $[S_{p_{k,j}^{(1)}}, L_{q_{k,j}^{(1)}}]$ and $[S_{p_{k,j}^{(2)}}, L_{q_{k,j}^{(2)}}]$: | Section 1.2 of main text |
| $\sigma_L$ and $\sigma$: multiplication factor to SS count | Section 1.2 of main text |
| $F$: objective function to be minimized | Section 1.2 of main text |
| $\sigma_L^{min}$ and $\sigma^{min}$: multiplication factor of minimize $F$ | Section 1.2 of main text |
| GA database | Section 2.1 of main text |
| GA block | Section 2.1 of main text |
| central GA zone | Section 2.1 of main text |
| inside, face, side, or corner GA block | Section 2.1 of main text (see Fig. 5) |
| $GA_j$: the $j$-th GA block | Section 2.1 of main text (see Fig. 6) |
| $\Lambda$: relative position of a zone in a block | Section 2.1 of main text |
| $[S, \Lambda]$ | Section 2.1 and section 2.2 of main text |
| $c^{GA_i}(S, \Lambda)$ | Section 2.1 of main text |
| $N_{surr}$ | Section 2.2 of main text |
| recovery zone (REC zone) | Section 2.2 of main text |
| recovery block (REC block) | Section 2.2 of main text |
| $REC_i$: the $i$-th REC block | Section 2.2 of main text |
| inside, face, side, and corner REC blocks | Section 2.2 of main text (see Fig. 5) |
| $c^{REC_i}(S, \Lambda)$ | Section 2.3 of main text |
| $c_{cor}(REC_i, GA_j)$: correlation coefficient between $REC_i$ and $GA_j$ | Section 2.4 of main text |
| $E_{simi}(REC_i, GA_j)$: score function | Section 2.4 of main text |
| $P(\sigma_L, \sigma_{L'})$ | Section 2.5 of main text |
| $E_{phys}$ | Section 2.5 of main text |
| $Gen_k$: the $k$-th generation | Sections 2.6 and 2.7 of main text (see Fig. 7) |
| $E_{score}(REC_i, GA_j)$ | Section 2.6 of main text |
| $N_{GA}^{all}$: number of members in a generation | Section 2.6 of main text |
| $N_{GA}^{top}$ | Section 2.7 of main text (see Fig. 7) |
| $N_{GA}^{mut}$ | Section 2.7 of main text |
| $N_{GA}^{exc}$ | Section 2.7 of main text |
| $MEM_k^{(1)}, MEM_k^{(2)}, ..., MEM_k^{(N_{GA}^{mut})}$ | Section 2.7 of main text |
| $GA^{(1)}, GA^{(2)}, ..., GA^{(N_{GA}^{mut})}$ | Section 2.7 of main text |
| mutation | Section 2.7 of main text |
| crossover/exchange | Section 2.7 of main text |
| $n_{snap}^{[1;M]}(L)$ | Section 4 of main text |

-----------------------------------------------------------------------------------------------------------------------------------------

## 1. Definition of reaction coordinates

Consider two atom groups $G_h^A$ and $G_h^B$ ($h = \alpha, \beta, \gamma, ...$) in a molecular system. A reaction coordinate (RC), $\lambda^{(h)}$, is defined by the distance between centers of mass of $G_h^A$ and $G_h^B$ (Fig. S1). Superscripts $A$ and $B$ indicate simply that the two atom groups are pairing to define $\lambda^{(h)}$, and then, one can exchange the superscripts as: $G_h^A \to G_h^B$ and $G_h^B \to G_h^A$ without changing the value of $\lambda^{(h)}$. The RC can be defined between two different molecules as in Fig. S1A or between two portions in a molecule (Fig. S1B). In Fig. S1C, the atom group $G_h^A$ is composed of two different portions in a molecule. As shown in this figure, there are various ways to define atom groups. This means that the current method is applicable to various problems (folding and binding).

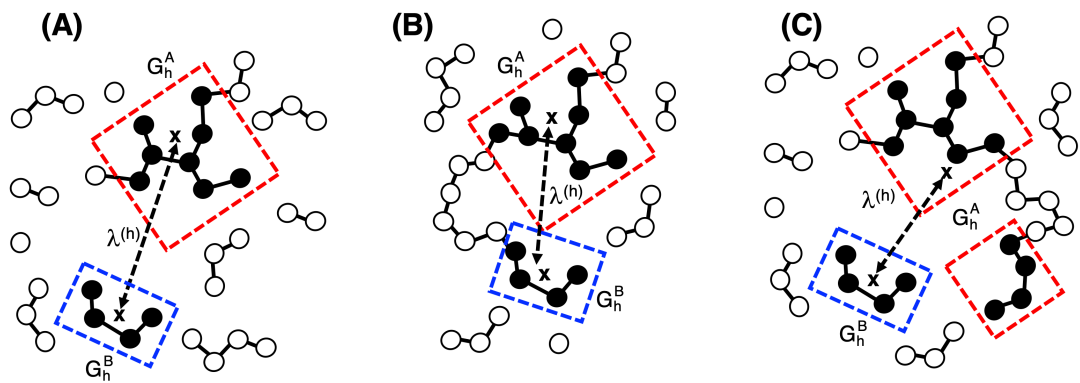

**Figure S1.** Two atom groups, $G_h^A$ and $G_h^B$, are indicated by red-colored and blue-colored rectangles, respectively. Atoms in $G_h^A$ and $G_h^B$ are presented by small black filled circles. Center of mass of each atom group is presented by a cross. The distance between the two centers of mass is $\lambda^{(h)}$ (broken-line with arrows).

## 2. Division of RC space and definition of virtual zones

In this section, we use a one-dimensional (1D) or two-dimensional (2D) RC space for explanation. Outcomes from the simplified RC space are extendable readily to those for a multi-dimensional (mD) RC space.

The RCs are arbitrary but should be computable uniquely from the system's conformation $\boldsymbol{R}$ as $\lambda^{(h)} = \lambda^{(h)}(\boldsymbol{R})$ ($h = \alpha, \beta, \gamma, ...$), where $\boldsymbol{R} = [x_1, y_1, z_1, x_2, y_2, z_2, \cdots]$ and $x_i$, $y_i$, and $z_i$ are the x-, y-, and z-coordinates of atom $i$, respectively. Then, we express the three RCs in a vector as $\boldsymbol{\lambda} = [\lambda^{(\alpha)}, \lambda^{(\beta)}, \lambda^{(\gamma)}, ...] = [\lambda^{(\alpha)}(\boldsymbol{R}), \lambda^{(\beta)}(\boldsymbol{R}), \lambda^{(\gamma)}(\boldsymbol{R}), ...]$. According to the motion of $\boldsymbol{R}$, $\boldsymbol{\lambda}$ moves in the mD RC space.

We divide each 1D RC axis into zones (called *RC zones* or simply *zones*) $\{\zeta_1^{(h)}, \cdots, \zeta_{n_{vs}(h)}^{(h)}\}$ as shown in Fig. S2A, where $\zeta_i^{(h)}$ is the $i$-th zone along the $\lambda^{(h)}$-axis ($i = 1, \cdots, n_{vs}(h)$), $n_{vs}(h)$ is the number of zones set to $\lambda^{(h)}$, and $\Delta\lambda_i^{(h)}$ is the width of the $i$-th zone. Usually we set $\Delta\lambda_i^{(h)} = const$ ($\forall i$), although $\Delta\lambda_i^{(h)}$ and $\Delta\lambda_j^{(h)}$ can be different for $i \neq j$ in general. Thus, we denote simply the width as $\Delta\lambda^{(h)}$ ($= \Delta\lambda_1^{(h)} = \cdots = \Delta\lambda_{n_{vs}(h)}^{(h)}$) in this paper. The case of $\Delta\lambda_i^{(h)} \neq \Delta\lambda_j^{(h)}$ is discussed in ref. 1. The lower and upper bounds of $\zeta_i^{(h)}$ are denoted as $[\zeta_i^{(h)}]_{min}$ and $[\zeta_i^{(h)}]_{max}$, respectively: $[\zeta_i^{(h)}]_{min} \leq \lambda^{(h)} \leq [\zeta_i^{(h)}]_{max}$ if $\lambda^{(h)} \in \zeta_i^{(h)}$. The entire sampling range for $\lambda^{(h)}$ is $[\zeta_1^{(h)}]_{min}$ to $[\zeta_{n_{vs}(h)}^{(h)}]_{max}$. Importantly, the adjacent zones $\zeta_i^{(h)}$ and $\zeta_{i\pm1}^{(h)}$ overlap to each other, whereas $\zeta_{i-1}^{(h)}$ and $\zeta_{i+1}^{(h)}$ do not. This overlap is essentially important for inter-zone transitions as explained later.

Figure S2B illustrates division of a 2D RC space constructed by $\lambda^{(\alpha)}$ and $\lambda^{(\beta)}$ into 2D zones. A 2D zone generated by the $i$-th and $j$-th 1D zones along $\lambda^{(\alpha)}$ and $\lambda^{(\beta)}$, respectively, is expressed as $[\zeta_i^{(\alpha)}, \zeta_j^{(\beta)}]$. For instance, the 2D zone enclosed by the broken line in Fig. S2B is expressed as $[\zeta_2^{(\alpha)}, \zeta_3^{(\beta)}]$. A higher dimensional RC space can be divided similarly: In the mD RC space constructed by $\lambda^{(\alpha)}$, $\lambda^{(\beta)}$, and $\lambda^{(\gamma)}$, a mD zone is expressed as $[\zeta_i^{(\alpha)}, \zeta_j^{(\beta)}, \zeta_k^{(\gamma)}, ...]$, where $\zeta_k^{(\gamma)}$ is the $k$-th 1D zone along $\lambda^{(\gamma)}$. Now, we introduce an index parameter $L^{(h)}$ ($h = \alpha, \beta, \gamma, ...$) to specify an RC zone along $\lambda^{(h)}$. A 2D index, $\boldsymbol{L} = [L^{(\alpha)}, L^{(\beta)}, L^{(\gamma)}, \ldots]$, is used to specify a 2D zone. In Fig. S2B, for instance, the index for the 2D zone enclosed by the broken line is presented as $\boldsymbol{L} = [2,3]$. Similarly, a mD index parameter is given as $\boldsymbol{L} = [L^{(\alpha)}, L^{(\beta)}, L^{(\gamma)}, \ldots]$.

Using the index parameter, a 2D zone is expressed as a vector: $\boldsymbol{\zeta_L} = \boldsymbol{\zeta}_{[L^{(\alpha)}, L^{(\beta)}]} = [\zeta^{(\alpha)}_{L^{(\alpha)}}, \zeta^{(\beta)}_{L^{(\beta)}}]$. Similarly, a mD zone is expressed as:

$$\boldsymbol{\zeta_L} = \boldsymbol{\zeta}_{[L^{(\alpha)}, L^{(\beta)}, L^{(\gamma)}, \ldots]} = [\zeta^{(\alpha)}_{L^{(\alpha)}}, \zeta^{(\beta)}_{L^{(\beta)}}, \zeta^{(\gamma)}_{L^{(\gamma)}}, \ldots], \tag{S1}$$

where $\zeta^{(h)}_{L^{(h)}}$ represents the $L^{(h)}$-th RC zone along $\lambda^{(h)}$. For instance, an index with $L^{(\alpha)} = i$, and $L^{(\beta)} = j$ in the 2D RC space is presented as $\boldsymbol{L} = [i, j]$, and the corresponding 2D zone is $\boldsymbol{\zeta_L} = \boldsymbol{\zeta}_{[i,j]} = [\zeta^{(\alpha)}_i, \zeta^{(\beta)}_j]$.

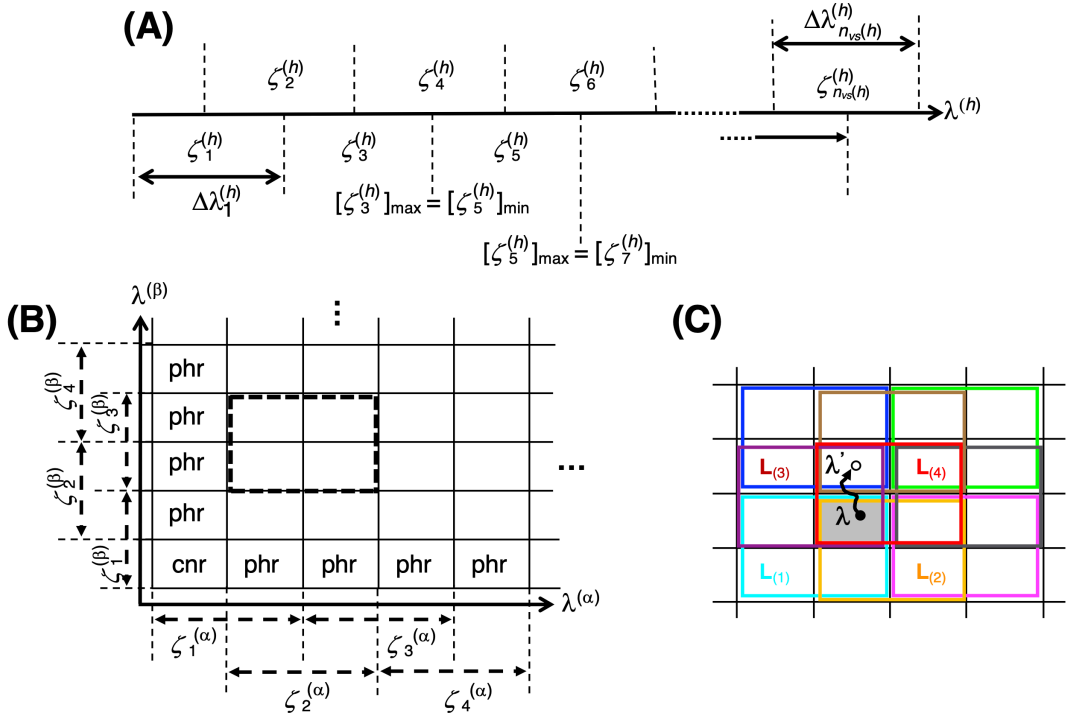

**Figure S2.** Schematic illustration of RC space. (A) Division of a 1D RC axis $\lambda^{(h)}$ ($h = \alpha, \beta, \gamma$) into RC zones $\{\zeta^{(h)}_i\}$ ($i = 1, \cdots, n_{vs}(h)$), where $\Delta\lambda^{(h)}_i$ is the width of the $i$-th zone, and $n_{vs}(h)$ is the number of zones along $\lambda^{(h)}$. The lower and upper bounds of the $i$-th zone are expressed respectively as $[\zeta^{(h)}_i]_{min}$ and $[\zeta^{(h)}_i]_{max}$. (B) Division of 2D RC space constructed by $\lambda^{(\alpha)}$ and $\lambda^{(\beta)}$ into 2D zones.

Periphery and corner regions are designated respectively as "phr" and "cnr". Broken-line frame is mentioned in text. (C) Close-up of 2D RC space, where four zones of differently colored frames overlap in the shaded region, which is called an intersection. System's conformation is moving from $\lambda$ to $\lambda'$. Colored indices $\boldsymbol{L}_{(1)}$, $\boldsymbol{L}_{(2)}$, $\boldsymbol{L}_{(3)}$, and $\boldsymbol{L}_{(4)}$ are mentioned in text. A color used for a character and frame correspond to each other. For instance, $\boldsymbol{L}_{(4)}$ specifies the red-colored zone. Shaded region is RC-zone intersection.

### 3. Potential energy and mD-VcMD

We set walls at the zone boundaries to confine $\lambda$ in a zone and allow inter-zone transitions occasionally during a simulation. Figure 2C is a portion of the 2D RC space, where the current zone is surrounded by red-colored frame. The shaded region (called *intersection*) in Fig. 2C is shared by four zones, which are referred to as *linked RC zones*, and expressed as $\zeta_{L_{(1)}}$ (cyan-colored frame), $\zeta_{L_{(2)}}$ (bright yellow), $\zeta_{L_{(3)}}$ (brown), and $\zeta_{L_{(4)}}$ (red). The indices $\boldsymbol{L}_{(1)}$, $\boldsymbol{L}_{(2)}$, $\boldsymbol{L}_{(3)}$, and $\boldsymbol{L}_{(4)}$ are those for the linked zones. Later, we renamed $\{\boldsymbol{L}_{(i)}\}$ ($i = 1, ...$) as linked virtual states. In general, we express the linked zones sharing an intersection as: $\boldsymbol{L}_{(1)}, ..., \boldsymbol{L}_{(n_{link})}$, where $n_{link}$ is the number of the linked RC zones for the intersection. The value of $n_{link}$ depends on the RC-space dimensionality and the position of the intersection in the RC space. For instance, in Fig. 2C, $n_{link} = 4$ if the intersection is inside of the 2D RC space, although $n_{link} = 2$ for intersections at periphery ("phr" in Fig. 2B), and $n_{link} = 1$ for intersections at corner ("cnr"). For a higher-dimensional RC space, $n_{link}$ increases. If the current RC zone is $\zeta_{L_{(4)}}$ in Fig. 2C, then $\lambda$ moves within the red-colored frame. Members of the linked RC zones change according to the motion: When $\lambda$ moves to $\lambda'$ in $\zeta_{L_{(4)}}$, then $\zeta_{L_{(1)}}$ and $\zeta_{L_{(2)}}$ are eliminated from the linked RC zones, and other two zones (blue- and green-colored framed zones) join in the linked RC zones.

We defined the potential energy $E_{entire}$ of the system as

$$E_{entire}(\boldsymbol{R}, \boldsymbol{L}) = E_R(\boldsymbol{R}) + E_V(\boldsymbol{L}) + E_{RV}(\lambda(\boldsymbol{R}), \boldsymbol{L}), \qquad (S2)$$

where $\boldsymbol{R}$ and $\boldsymbol{L}$ are dynamic variables to specify $E_{entire}$. Therefore, we rename the current-zone index $\boldsymbol{L}$ a *virtual-state variable* (or simply, *virtual state*) to express that the system's motion is controlled by varying $\boldsymbol{L}$. The term $E_R(\boldsymbol{R})$ is the original potential energy described by the system's conformation $\boldsymbol{R}$, $E_{RV}(\lambda, \boldsymbol{L})$ is introduced to confine $\lambda$ in the current RC zone $\zeta_L$, and $E_V(L)$ controls a transition of the current zone from $\zeta_L$ to $\zeta_{L'}$.

The form of $E_{RV}$ is a well-type potential expressed as:

$$E_{RV}(\lambda; L) = \begin{cases} 0 & (\text{for } \lambda \in \zeta_L) \\ c_{RV} \sum_h \left(\delta_{L^{(h)}}^{(h)}\right)^2 & (\text{for } \lambda \notin \zeta_L) \end{cases} \quad (S3)$$

where

$$\delta_{L^{(h)}}^{(h)} = \begin{cases} \lambda^{(h)} - [\zeta_{L^{(h)}}^{(h)}]_{min} & (\text{for } \lambda^{(h)} < [\zeta_{L^{(h)}}^{(h)}]_{min}) \\ \lambda^{(h)} - [\zeta_{L^{(h)}}^{(h)}]_{max} & (\text{for } \lambda^{(h)} > [\zeta_{L^{(h)}}^{(h)}]_{max}) \end{cases}. \quad (S4)$$

The summation for $h$ in Eq. S3 is taken over $\alpha$, $\beta$, $\gamma$,…. Remember that $[\zeta_{L^{(h)}}^{(h)}]_{min}$ and $[\zeta_{L^{(h)}}^{(h)}]_{max}$ are the lower and upper boundaries of the $L^{(h)}$-th RC zone along the $\lambda^{(h)}$ axis, respectively. The restoring force, $-\nabla E_{RV}$, acts on $\lambda$ only when $\lambda$ is outside the zone $\zeta_L$, and the coefficient $c_{RV}$ is set to a large value so that the restoring force increases rapidly outside $\zeta_L$. Usually we set $c_{RV}$ to 100 kcal/mol. We count the number of snapshots detected in each RC zone. However, we eliminate the outside-the-zone conformations from the count. Therefore, the effect of the artificial force (i.e., restoring force) is removed completely in the statistical analyses.

We define $E_V$ as follows:

$$E_V(L) = g_L, \quad (S5)$$

where $g_L$ is a constant in $\zeta_L$ independent of $\lambda$ and $R$, but varies when $L$ moves to another zone $L'$: $g_L \neq g_{L'}$ if $L \neq L'$. Because $L$ varies discretely, we use a Monte-Carlo method to move $L$, which means that $E_V$ is related to transition probabilities among virtual states.

**4. Inter-virtual state transition probability**

When an inter-zone transition is applied to the system at a time $t$ ($= t_0 + n \Delta t$), where $t_0$ is the initial time of simulation and $n$ is an integer. In other words, $R$ is confined in a current zone $\zeta_L$ for a time interval of $[t_0 + (n-1)\Delta t; t_0 + n\Delta t]$. Now, we explain the simulation procedure concretely: In a time interval $[t_0, t_0 + \Delta\tau]$, the *configurational motion* (CFM); the motion of $\boldsymbol{R}$) is performed with an MD protocol where forces acting on atoms are derived by $-\nabla E_{entire}$ with fixing the current virtual state at $\boldsymbol{L}$ (i.e., the current RC zone is $\zeta_L$). Suppose that $\boldsymbol{\lambda}$ is in an intersection shared by $n_{link}$ linked RC zones $\zeta_{L_{(1)}}$, $\zeta_{L_{(2)}}$,…, and $\zeta_{L_{(n_{link})}}$ at time $t_0 + \Delta\tau$, and note that the current RC zone $\zeta_L$ is one of the $n_{link}$ linked RC zones. The virtual states whose zones are linked are called *linked virtual states*. Then, *inter-virtual state transition* (IVT) from $\boldsymbol{L}$ to $\boldsymbol{L}'$ (i.e., $\boldsymbol{L}'$ is also one of $\boldsymbol{L}_{(1)}$,…, $\boldsymbol{L}_{(n_{link})}$) is executed at time $t_0 + \Delta\tau$ by selecting one of the $n_{link}$ linked RC zones with using a transition probability between $\boldsymbol{L}$ and $\boldsymbol{L}'$. If $\zeta_{L'} = \zeta_L$, this transition is a self-transition. Note that $\boldsymbol{R}$ and $\boldsymbol{\lambda}$ are fixed when executing IVT. In the next interval $[t_0 + \Delta\tau, t_0 + 2\Delta\tau]$, CFM is performed using $E_{entire}$ with fixing $\boldsymbol{L}'$ (i.e., $\boldsymbol{\lambda}$ moves in $\zeta_{L'}$). Then, at time $t_0 + 2\Delta\tau$, IVT is achieved from $\boldsymbol{L}'$ to $\boldsymbol{L}''$ with fixing $\boldsymbol{R}$ and $\boldsymbol{\lambda}$, and so on. Although the interval $\Delta\tau$ can be set arbitrary in theory, we usually set $\Delta\tau$ to a constant value (say 10 ps).

Needless to say, the transition probability controls the sampling efficiency. Here, we present an optimal set of the transition probability. First, the formal expression for the transition probability is given as:

$$P_t(\boldsymbol{L} \to \boldsymbol{L}') = exp[-\beta\Delta H_{entire}]$$
$$= exp[-\beta(H_{entire}(\boldsymbol{R}, \boldsymbol{L}') - H_{entire}(\boldsymbol{R}, \boldsymbol{L}))]$$
$$= exp[-\beta\Delta E_V] \qquad (S6)$$

where $P_t(\boldsymbol{L} \to \boldsymbol{L}')$ is the *inter-virtual state transition probability* (*IVT probability*) from $\boldsymbol{L}$ to $\boldsymbol{L}'$ with fixing $\boldsymbol{R}$ and $\boldsymbol{\lambda}$ at $t$, and $\Delta E_V = E_V(\boldsymbol{L}') - E_V(\boldsymbol{L})$. Note that $\zeta_L$ and $\zeta_{L'}$ are zones sharing an intersection, which involves $\boldsymbol{\lambda}$. The terms $E_R(\boldsymbol{R})$ and $E_{RV}(\boldsymbol{\lambda}, \boldsymbol{L})$ are canceled out in Eq. S6: $\Delta E_R = \Delta E_{RV} = 0$. Therefore, $P_t(\boldsymbol{L} \to \boldsymbol{L}')$ is controlled only by $E_V$. The term "inter-zone transition" used in the above sections is replaced by "inter-virtual state transition" here. Equation 6 is a formal expression for IVT transitions, where $E_V$ (or $g_L$) is not derived from a physicochemical law but can be set according to our purpose. This is why we can modulate $P_t(\boldsymbol{L} \to \boldsymbol{L}')$ to increase sampling efficiency.

We introduce a quantity named *virtual state-partitioned probability* $Q_{entire}(L)$, which is the number of snapshots detected in the zone $\zeta_L$ in the simulation. Suppose that the system is at $\lambda$ in an intersection at a time and that its virtual state is $L_{(i)}$, which is one of linked virtual zones $\{L_{(1)}, ..., L_{(n_{link})}\}$. Then, one count of detection is added to $Q_{entire}(L_{(i)})$ not to those of the other linked virtual states. Furthermore, we introduce another quantity $Q_{cano}(L)$, called a *virtual state-partitioned canonical probability* [1,2]. If we perform a long mD-VcMD simulation with setting all IVT probabilities to a constant (i.e., $P_t(L \to L') = const$ for $\forall L$ and $\forall L'$), then the resultant $Q_{entire}(L)$ is $Q_{cano}(L)$. Thus, $Q_{cano}(L)$ is a special case of $Q_{entire}(L)$. However, such a simulation takes considerably long computation time for convergence of $Q_{cano}(L)$. Last in this paragraph, we note that $Q_{entire}(L)$ and $Q_{cano}(L)$ are not continuous but discrete function of $L$ because the zone index is a discrete quantity.

Determination of $Q_{cano}(L)$ is essentially important because we can define the IVT probability using $Q_{cano}(L)$. Ref. [3] showed that sampling enhancement is realized when $Q_{entire}(L) \approx const$ is obtained during sampling, and finally the IVT probability is given as:

$$J_{L_{(i)}} = \left[ Q_{cano}(L_{(i)}) \sum_{m=1}^{n_{link}} \frac{1}{Q_{cano}(L_{(m)})} \right]^{-1}. \tag{S7}$$

In general, the current virtual zone is one of linked zones $\zeta_{L_{(1)}}, \zeta_{L_{(2)}}, ..., \zeta_{L_{(n_{link})}}$. Then, $J_{L_{(i)}}$ provides an IVT probability from any of the linked virtual states to $L_{(i)}$. If $\lambda$ is in the corner zone (see Fig. 2B), $n_{link}$ may be 1, and then, the only self-transition is allowed. We note that Eq. S6 is a formal expression for the IVT probability and Eq. S7 is the actually used one to optimize the sampling.

One may consider that another expression, $E_V(L) = -RT \ln[Q_{cano}(L)]$, can result in $Q_{entire}(L) \approx const$. Equation S7 is another expression to obtain $Q_{entire}(L) \approx const$ with a small rejection rate in IVT [3].

Whereas Eq. 7 requires a set of $Q_{cano}(L)$, it is difficult to obtain accurate $Q_{cano}(L)$ directly from a simulation with setting $P_t(L \to L') = 1$ in a realistic time. Instead, we use a method to convert $Q_{entire}(L)$ to $Q_{cano}(L)$ via iterative mD-VcMD simulations. Section 1 of SI explains briefly the "original mD-VcMD method"[1,2] for the conversion. Then, below we present two methods: "a subzone-based mD-VcMD method", which is an extension of the original method, and a "GA-guided mD-VcMD method", which is introduced to expand sampling to un-sample regions through iterative simulation.

## 5. Original method to compute $Q_{cano}(L)$

Assume that we have performed the $M$-th iteration of mD-VcMD. Then, $Q_{cano}(L)$ is set as [1,2]:

$$Q_{cano}^{[M]}(L) = Q_{cano}^{[M-1]}(L) Q_{entire}^{[M]}(L), \qquad (S8)$$

where $Q_{cano}^{[M-1]}(L)$ is $Q_{cano}(L)$ obtained from the $(M-1)$-th iteration, and $Q_{entire}^{[M]}(L)$ is the resultant $Q_{entire}(L)$ calculated numerically from the $M$-th iteration (i.e., the currently done simulation). From $Q_{cano}^{[M]}(L)$ computed by Eq. S8, $J_{L^{(i)}}$ is calculated (Eq. S7), and the $(M+1)$-th iteration is performed. This iteration is continued till a relation $Q_{entire}^{[M]}(L) \approx const$ ($\forall L$) is satisfied approximately [1,2]. Because $Q_{cano}^{[0]}(L)$ is not prepared in advance for the first iteration, we assume $Q_{cano}^{[0]}(L) = const$ ($\forall L$) and $Q_{cano}^{[1]}(L) = Q_{entire}^{[1]}(L)$ as a guess.

**Figure S3. Flow chart of the GA-guided mD-VcMD method.**

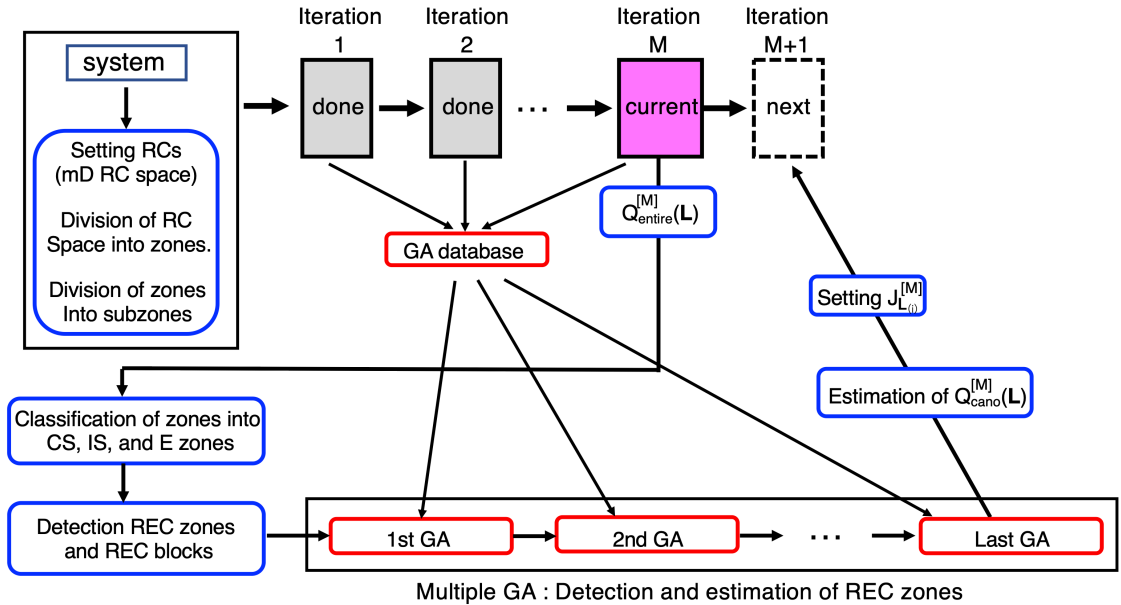

**Figure S3.** Schematic illustration of GA-guided mD-VcMD. Given a system, RCs are set, and the multi-dimensional (mD) RC space is divided into zones. Then, the zones are divided into subzones. Next,

iteration $M$ (current iteration) is performed, and $Q_{entire}^{[M]}(L)$ is obtained. After that, GA blocks are generated using data from iterations 1 to $M$. The ensemble of GA blocks is GA database. Sampled zones from iteration $M$ are classified into three: CA, IC, and E zones. Then, REC zones and REC blocks are detected from the patterns of spatial patterns of CA, IC, and E zones. Next, the first GA procedure is done with comparing the spatial patterns of the SS counts between REC blocks and GA blocks, from which SS counts of the REC zones are determined. Then, new REC zones are detected, and the second GA procedure is done to determine the new REC zones, and so on. When the SS counts of all REC zones are determined, the multiple GA procedure is finished. Resultant SS counts are used to calculate $Q_{entire}^{[M]}(L)$, from which IVT probability $J_{L_{(i)}}^{[M]}$ is calculated. Iteration $M + 1$ is performed using $J_{L_{(i)}}^{[M]}$, and so on.

**References**

[1] Higo, J., Kasahara, K., Wada, M., Dasgupta, B., Kamiya, N., Hayami, T., *et al.* Free-energy landscape of molecular interactions between endothelin 1 and human endothelin type B receptor: Fly-casting mechanism. *Protein Engineering, Design & Selection* (*PESD*) **32**, 297–308 (2019). DOI: 10.1093/protein/gzz029

[2] Hayami, T., Higo, J., Nakamura, H. & Kasahara, K. Multidimensional virtual-system coupled canonical molecular dynamics to compute free-energy landscapes of peptide multimer assembly. *J Comput. Chem.* **40**, 2453–2463 (2019). DOI: 10.1002/jcc.26020

[3] Higo, J., Kasahara, K., Dasgupta, B. & Nakamura, H. Enhancement of canonical sampling by virtual-state transitions. *J. Chem. Phys.* **146**, 044104 (2017). DOI: 10.1063/1.4974087